\documentclass{article}
\usepackage{arxiv}
\usepackage{amsthm} 
\usepackage{amssymb}
\usepackage{amsmath}
\usepackage{empheq}
\usepackage{amscd}
\usepackage{graphicx}
\usepackage{graphics}
\usepackage[noadjust]{cite}
\usepackage{amsthm}
\usepackage{caption}
\usepackage{multirow}
\usepackage{array}
\usepackage{rotating}
\usepackage{hyperref}

\usepackage{indentfirst}
\usepackage{tikz}
\usepackage{calc}
\usepackage{epsfig}
\usepackage[numbers]{natbib}
\usepackage[makeroom]{cancel}
\usepackage{multicol}
\usepackage{csquotes}
\usepackage{enumitem}

\usepackage{hyperref}       
\hypersetup{colorlinks,linkcolor={green},citecolor={green},urlcolor={black}}
\usepackage{optidef}

\renewcommand{\S}[0]{\bold{U}}                   
\newcommand{\G}[0]{\bold{S}}                   

\newcommand{\eref}[1]{equation (\ref{#1})}                  
\DeclareMathOperator*{\minimize}{minimize}

\begin{document}
\title{An extended Gauss-Newton method for full waveform inversion}

\author{\href{https://orcid.org/0000-0002-9879-2944}{\hspace{1mm}Ali Gholami} \\
  Institute of Geophysics, Polish Academy of Sciences, Warsaw, Poland\\
  \texttt{agholami@igf.edu.pl} \\ 
  }

\graphicspath{{./figs/}}

\renewcommand{\shorttitle}{Extended Gauss-Newton method ~~~~~~~~~~~~~~~~~~ Ali Gholami}

\maketitle

\begin{abstract}
Full waveform inversion (FWI) is a large-scale nonlinear ill-posed problem for which computationally expensive Newton-type methods can become trapped in undesirable local minima, particularly when the initial model lacks a low-wavenumber component and the recorded data lacks low-frequency content. A modification to the Gauss-Newton (GN) method is proposed to address these issues. The standard GN system for multisource multireceiver FWI is reformulated into an equivalent matrix equation form, with the solution becoming a diagonal matrix rather than a vector as in the standard system. The search direction is transformed from a vector to a matrix by relaxing the diagonality constraint, effectively adding a degree of freedom to the subsurface offset axis. The relaxed system can be explicitly solved with only the inversion of two small matrices that deblur the data residual matrix along the source and receiver dimensions, which simplifies the inversion of the Hessian matrix. When used to solve the extended source FWI objective function, the Extended GN (EGN) method integrates the benefits of both model and source extension.  The EGN method effectively combines the computational effectiveness of the reduced FWI method with the robustness characteristics of extended formulations and offers a promising solution for addressing the challenges of FWI. It bridges the gap between these extended formulations and the reduced FWI method, enhancing inversion robustness while maintaining computational efficiency. The robustness and stability of the EGN algorithm for waveform inversion are demonstrated numerically.
\end{abstract}
\section{Introduction}
\noindent Seismic waveform inversion is a nonlinear optimization problem that aims to estimate the physical properties of the Earth from recorded seismic data. Traditionally, the best estimate is obtained by minimizing the squared two norm distance between the computed synthetic seismograms and the recorded data \citep[see, e.g.,][]{Tarantola_1984_ISR,Pratt_1998_GNF,Virieux_2009_OFW}. Iterative methods, such as Newton-type methods \citep{Pratt_1998_GNF,Metivier_2017_TRU}, are commonly used to calculate this estimate. The crucial step in these methods is to compute a direction in which a new estimate can be searched, resulting in a better fit to the data. This direction is determined by solving the Newton system. However, the success of these methods heavily relies on accurately solving this system at each iteration to estimate a suitable model update.

The preconditioned steepest-descent (PSD) method is a popular algorithm for full waveform inversion (FWI) due to its computational efficiency, as it avoids computing the Hessian matrix \citep{Gauthier_1986_TDN}. Its main requirement is to compute the gradient of the misfit function with respect to the model parameters (i.e., squared slowness). This can be accomplished efficiently by employing the first-order adjoint-state method, which involves performing zero-lag cross-correlation between the forward-propagated source-side wavefield and backward-propagated receiver-side wavefield, with the data residual for the background model serving as the adjoint source generating the receiver wavefield \citep{Plessix_2006_RAS}. However, its convergence rate is relatively slow, necessitating a large number of iterations to reach a stationary point. To address this issue, truncated versions of the Newton method have been proposed, which implement the Hessian matrix through a finite number of conjugate gradient (CG) iterations \citep{Metivier_2017_TRU}. In this approach, the Hessian vector products, which are required at each CG iteration, are performed implicitly using a matrix-free formalism, such as second-order adjoint state formulas. This reduces the computational load by introducing new wave equations that must be solved for each CG iteration. 
However, the singular or ill-conditioned nature of the Hessian matrix can make determining an appropriate search direction difficult, especially when the initial model is far from the global solution. Overcoming these challenges and improving the efficacy and robustness of FWI methods are ongoing research topics in the field.

Several algorithms have been proposed in recent decades to improve the performance of FWI. In this paper, we will look at some relevant methods that aim to address the shortcomings of traditional approaches. One method is to expand the optimization problem by introducing artificial parameters. These parameters are penalized during the velocity model update, ensuring that the original problem is reached at the convergence point \citep{Symes_2008_MVA}. Model extension and source extension are two popular approaches to extension.

FWI based on model extension works by splitting the model into short-wavelength (perturbation) and long-wavelength (background) components. \citet{Symes_1994_IRSb}, for example, employed separate perturbation models for each source, extending the model along the surface offset or source coordinate.  
Similarly, the model can be extended along the subsurface offset, which corresponds to the distance between two subsurface points in the extended Born equation related to the incident and reflected wavefields \citep{Claerbout_1985_IEI,Rickett_2002_OAC}. It is also possible to extend along the scattering angle axis  \citep{Sava_2003_ADC} or the time-lag axis \citep{Biondi_2014_SIF}.
Compared to the standard FWI, these methods are more robust but their computational and memory costs can be high due to the increased dimensionality of the problem \citep{Biondi_2014_SIF}.

In the extended-source FWI, the optimization is performed over the source extension term \citep{Huang_2018_VSE} or the extended wavefield \citep{VanLeeuwen_2013_MLM}, in addition to the model parameters, and the algorithm penalizes the extending parameter while updating the velocity, ensuring that the physics is satisfied at the convergence point. This is done in an alternating optimization framework that solves the constrained FWI problem with a quadratic penalty formulation.  These methods have been demonstrated to be more robust than the conventional FWI, but they have high computational and memory costs. To address this issue,  \citet{vanLeeuwen_2016_PMP} eliminated the wavefield variable, resulting in a modified objective function that only included the model parameters. This modified function was later shown to be equivalent to an objective function in the standard FWI form with a model-dependent weighting matrix \citep{vanLeeuwen_2019_ANO,Symes_2020_WRI,Gholami_2022_EFW}.
The gradient remains very similar to that of the standard FWI in this formulation; it is the zero-lag cross-correlation between extended forward and backward wavefields.  The former is calculated by solving an augmented wave equation and the latter is calculated by back propagating an extended data residual. The FWI data residual is deblurred along the receiver axis with a blurring matrix that is the data-space Hessian of the receiver-side Green functions to obtain this extended data residual \citep{Gholami_2022_EFW}. The associated GN Hessian is also very similar to that of the standard FWI, except for the extra weighting matrix and the replacement of the standard reduced wavefield with the extended one \citep[][ their equation 25]{Gholami_2022_EFW}. In practice, the Hessian is replaced with a sparse approximation, hence performing a PSD optimization \citep{vanLeeuwen_2016_PMP}.

In this paper, we propose an alternative extended algorithm for FWI by modifying the GN formulation. We begin with the standard squared two norm misfit function for frequency-domain multisource multireceiver FWI and solve it using the GN method. The GN search direction is a vector that solves the GN system in its standard form. We show that the GN system can be expressed as a matrix equation using the relationship between matrix multiplication and the Hadamard (element-wise) product operation \citep{Horn_1994_TMA}, with the solution being a diagonal matrix with the search direction on its main diagonal. This constraint is relaxed by allowing the search direction to be a matrix rather than a vector. This relaxation entails introducing an extra degree of freedom in the subsurface offset axis for the search direction, effectively sampling the subsurface offset space in terms of range and orientation \citep[including horizontal and non-horizontal offsets, see e.g.,][]{Biondi_2002_PIO,Biondi_2004_ADC,Symes_2008_MVA}. More importantly, it leads to a separable Hessian matrix in which the source and receiver parts are properly factored, allowing the relaxed system to be explicitly solved. This only necessitates the inversion of two separate Hessian matrices of the size of the number of sources and the number of receivers, which are applied along the source and receiver coordinates of the FWI data residual matrix, respectively. The receiver side Hessian is the same as in the source extended FWI discussed above. The deblurring effect of the source-side Hessian is similar along the source coordinate. As a result, unlike the extended source method, which uses a 1D deblurring filter on the data residual along the receiver coordinate, the extended GN (EGN) method uses a 2D kernel along the source and receiver coordinates to deblur the data residuals. 

The proposed EGN approach can also be used for solving the (reduced) penalty objective function of extended-source FWI \citep{vanLeeuwen_2016_PMP}, offering potential advantages in terms of computational efficiency and robustness. By replacing the standard reduced wavefields with extended wavefields, the EGN method establishes a connection between extended waveform inversion based on model extension and those based on source extension.
In this case, the proposed EGN algorithm lies at the interface between standard reduced FWI, extended-source FWI, and extended-model FWI. By incorporating the extended search direction and leveraging the separability of the Hessian matrix, it provides a promising approach for addressing the FWI problem. The explicit solution of the relaxed system and the potential computational benefits make the EGN algorithm an intriguing method to explore for enhancing the efficiency and effectiveness of FWI.

The remainder of the paper is organized as follows: we provide the necessary notation and preliminaries that are not the main focus of the paper, followed by the introduction of the EGN theory. We also include details on computing the reduced search direction from the extended one, selecting the step length, increasing efficiency through random projections, and discussing the computational complexity of EGN. Subsequently, we apply EGN to solve the extended-source FWI. We present three sets of numerical examples to illustrate the effectiveness of EGN. Finally, we conclude with discussions and conclusions.

\section{Notation and Preliminaries}
\noindent In this section, we introduce important notation, definitions, and fundamental matrix identities that are utilized throughout the paper. Vectors and matrices are denoted by bold lowercase and uppercase letters, respectively. The transpose and conjugate transpose of a matrix are represented by the superscript $t$ and $T$, respectively. The complex conjugate of a quantity is denoted by an overline. The Hadamard multiplication is denoted by the symbol $\circ$. The angular frequency is denoted as $\omega$, the Laplacian as $\boldsymbol{\Delta}$, and the gradient operator as $\nabla$.

For column $N$-vectors $\bold{u}$ and $\bold{v}$, $\langle \bold{u},\bold{v}\rangle=\overline{\langle \bold{v},\bold{u}\rangle}=\bold{u}^T\bold{v}$ represents their inner product, and $\|\bold{u}\|_2=\sqrt{\langle\bold{u},\bold{u}\rangle}$ denotes the two norm of the vector.
For $M\times N$ matrices $\bold{A}$ and $\bold{B}$, $\langle \bold{A},\bold{B}\rangle_F=\overline{\langle \bold{B},\bold{A}\rangle_F}=\text{trace}(\bold{A}^T\bold{B})$ represents the Frobenius inner product, and $\|\bold{A}\|_F=\sqrt{\langle \bold{A},\bold{A} \rangle_F}$ denotes the Frobenius norm of the matrix. For a scalar function $E(\bold{u})$, the gradient $\nabla E(\bold{u})$ is defined as a column vector that contains the partial derivatives of $E$ with respect to each component of $\bold{u}$. For a vector function $\bold{F}(\bold{u})$, the Jacobian $\bold{J}=\nabla \bold{F}(\bold{u})$ is a matrix whose rows consist of the gradients of the vector components with respect to $\bold{u}$.
Additionally, $\text{diag}(\bold{u})$ represents a diagonal matrix with the vector $\bold{u}$ along its main diagonal. These notations and definitions are employed throughout the paper to facilitate the understanding of subsequent derivations and discussions.

%
Consider $\bold{A}$ and $\bold{B}$ as $M\times N$ matrices, and $\bold{u}$ as an $N$-length vector. In this case, the $i$th diagonal element of the matrix $\bold{C}=\bold{A}\text{diag}(\bold{u})\bold{B}^t$ is equal to the $i$th entry of the vector $(\bold{A}\circ \bold{B})\bold{u}$. This relationship can be expressed as follows:
\begin{equation} \label{Hadamard_diag}
\bold{C}_{ii}=\left[\bold{A}\text{diag}(\bold{u})\bold{B}^t\right]_{ii}=
[(\bold{A}\circ \bold{B})\bold{u}]_i,
\end{equation}
where $\bold{C}_{ii}$ represents the $i$th diagonal element of the matrix $\bold{C}$ \citep[see page 305 of][]{Horn_1994_TMA}. This equation demonstrates the correspondence between the $i$th diagonal entry of the matrix $\bold{C}$ and the $i$th entry of the vector resulting from the Hadamard product of $\bold{A}$ and $\bold{B}$, multiplied by $\bold{u}$.

Consider $\bold{C}$ and $\bold{D}$ as square invertible matrices of size $M\times M$ and $N\times N$, respectively, and $\bold{U}$ and $\bold{V}$ as two rectangular matrices of size $M\times N$ and $N\times M$ such that $\left(\bold{D} - \bold{V}\bold{C}^{-1}\bold{U}\right)$ and $\left(\bold{C} - \bold{U}\bold{D}^{-1}\bold{V}\right)$ are invertible. In this case, the following matrix identities hold \citep{Guttman_1946_EMC}:
\begin{equation} \label{Guttman_identity}
\left(\bold{D} - \bold{V}\bold{C}^{-1}\bold{U}\right)^{-1}\bold{VC}^{-1}=
\bold{D}^{-1}\bold{V}\left(\bold{C} - \bold{U}\bold{D}^{-1}\bold{V}\right)^{-1},
\end{equation}
\begin{equation} \label{Guttman_identity_sub}
\left(\beta\bold{I} + \bold{V}\bold{U}\right)^{-1}\bold{V}=
\bold{V}\left(\beta\bold{I} + \bold{U}\bold{V}\right)^{-1},
\end{equation}
where $\beta>0$, and $\bold{I}$ denotes the identity matrix of an appropriate size. 
These identities, in fact, provide efficient methods for solving normal equations involving the matrices $\bold{C}$, $\bold{D}$, $\bold{U}$, and $\bold{V}$. By utilizing these identities, one can simplify computation and achieve more efficient solutions for problems requiring the inversion or manipulation of rectangular matrices. 

\section{Theory}
\noindent 
The frequency-domain FWI can be formulated as a least-squares problem, aiming to minimize the misfit between the observed data and the data predicted by the forward operator. The objective function in this formulation is given by 
\citep{Pratt_1998_GNF}: 
\begin{equation} \label{main_obj}
\minimize_{\bold{m}}E(\bold{m})=\frac12\sum_{s=1}^{N_s}\|\bold{F}_{\!\!s}(\bold{m})-\bold{d}_s\|_2^2,
\end{equation}
where $\bold{m}$ represents the model parameters to be estimated. The forward operator $\bold{F}_{\!\!s}(\bold{m})$ maps the model space to the data space, and it is defined as
\begin{equation} \label{F}
\bold{F}_{\!\!s}(\bold{m}) = \bold{P}\bold{A(m)}^{-1}\bold{b}_s.
\end{equation}
Here, $s$ denotes the source index, $N_s$ is the number of sources, and $N_r$ is the number of receivers.
The observed data $\bold{d}_s$ and the predicted data $\bold{F}_{\!\!s}(\bold{m})$ are complex-valued vectors of size $N_r\times 1$. The forward operator involves the sampling operator $\bold{P}$, which samples the wavefield $\bold{u}_s$ at the receiver locations. The wavefield $\bold{u}_s$ is computed by solving the Helmholtz equation with the squared slowness $\bold{m}$, represented by the Helmholtz operator $\bold{A(m)}$. The Helmholtz operator consists of the Laplacian operator $\boldsymbol{\Delta}$ and the squared slowness term $\text{diag}(\bold{m})$ weighted by the square of the angular frequency: $\bold{A(m)} = \boldsymbol{\Delta} + \omega^2 \text{diag}(\bold{m})$.

Newton-type methods are commonly employed to solve the least squares problem in equation \ref{main_obj} by iteratively linearizing the objective function. Starting from an initial model $\bold{m}_0$, the iterative process generates a sequence of models $\{\bold{m}_k\}$ using the update equation:
\begin{equation}
\bold{m}_{k+1}=\bold{m}_k+\alpha_k \delta\bold{m}_k,
\end{equation}
where $\alpha_k$ denotes the step length and $\delta\bold{m}_k$ represents the search direction. In the subsequent explanation, we will assume that the dependency of the variables on the iteration number $k$ has been removed for the sake of compact notation. This means that the variables like $\bold{m}$ and $\delta\bold{m}$ are implicitly understood to represent their values at the current iteration without explicitly denoting the iteration index. This simplification allows for a more concise representation of the iterative process and equations.

In the Newton method, the search direction $\delta\bold{m}$ is determined by solving the Newton system given by 
\begin{equation} \label{Newtonsystem}
\bold{H}\delta\bold{m} = -\nabla E,
\end{equation}
where $\nabla E(\bold{m})$ is the gradient vector and $\bold{H}(\bold{m})=\nabla^2 E(\bold{m})$ is the Hessian matrix. The gradient vector is calculated by computing the partial derivatives of $E$ with respect to the model parameters, 
\begin{equation} \label{gradvec}
\nabla E(\bold{m})= \sum_{s=1}^{N_s}\bold{J}_{\!s}^T\delta\bold{d}_s.
\end{equation}
It involves the data residual vectors $\delta\bold{d}_s=\bold{F}_{\!\!s}(\bold{m})-\bold{d}_s$, which represent the difference between the predicted data $\bold{F}_{\!\!s}(\bold{m})$ and the observed data $\bold{d}_s$. The Jacobian matrices $\bold{J}_{\!s}(\bold{m})$, defined in equation \ref{J}, represent the gradients of the forward operator $\bold{F}_{\!\!s}(\bold{m})$ with respect to the model parameters.
\begin{equation} \label{J}
\bold{J}_{\!s}(\bold{m})=\nabla \bold{F}_{\!\!s}(\bold{m})=-\bold{PA}(\bold{m})^{-1}\frac{\partial \bold{A(m)}}{\partial \bold{m}}\bold{u}_s,
\end{equation}
where $\bold{A(m)}\bold{u}_s=\bold{b}_s$.
The Hessian matrix $\bold{H}(\bold{m})$ is computed by summing the contributions from first-order information provided by the Jacobian normal matrix and the nonlinear term $\bold{R}_s(\bold{m})$ that captures second-order information, i.e.
\begin{equation}
\bold{H}(\bold{m}) = \sum_{i=1}^{N_s}[\bold{J}_{\!s}(\bold{m})^T\bold{J}_{\!s}(\bold{m}) + \bold{R}_s(\bold{m})].
\end{equation}
It is extremely difficult to solve equation \ref{Newtonsystem} with this Hessian matrix because the nonlinear term may result in an indefinite Hessian unless the initial model is sufficiently close to the global minimum of the objective function \citep{Pratt_1998_GNF}. The GN method employs a positive semidefinite approximate Hessian obtained by ignoring the Hessian's nonlinear component. This approximation is used to solve the GN system, which determines the search direction. Traditional methods, such as the CG method, which introduces additional wave-equation solves per CG iteration, typically approximate the solution of this system. However, the issue of convergence to a local minimum persists \citep{Metivier_2017_TRU}.

In the subsection that follows, we reformulate the GN system by separating the effects of the shots and receivers in the gradient and Hessian using the Hadamard product. Then, in order to effectively solve the resulting system and benefit from the advantages of the model extension  \citep{Rickett_2002_OAC,Sava_2003_ADC,Symes_2008_MVA,Biondi_2014_SIF}, we extend the GN search direction.

\subsection{Extended Gauss-Newton method}
\noindent 
The GN search direction, denoted as $\delta\mathbf{m}$, is computed by solving the following equation:
\begin{equation} \label{Gauss_Newton}
\left[\sum_{s=1}^{N_s}\mathbf{J}_{\!s}^T\mathbf{J}_{\!s}\right] \delta\mathbf{m}=-\sum_{s=1}^{N_s}\mathbf{J}_{\!s}^T\delta\mathbf{d}_s.
\end{equation}
Here, $\mathbf{J}_{\!s}$ represents the Jacobian matrix associated with the $s$th source. To facilitate the notation, we introduce the source matrix $\mathbf{B}$ as:
\begin{equation}
\bold{B}=\begin{bmatrix} \bold{b}_1& \bold{b}_2&\cdots& \bold{b}_{N_s}\end{bmatrix}.
\end{equation}
Additionally, we define the wavefield matrix $\S$ and the receiver-side Green function matrix $\G$ as follows:
\begin{equation} \label{S_GN}
\S = \omega^2\bold{B}^T\bold{A}^{-T},\qquad \G=\bold{PA}^{-1}.
\end{equation}
Utilizing the definition of the Jacobian matrix, \eref{J}, we can express the gradient vector as:
\begin{align} \label{grad_Hadamard}
\sum_{s=1}^{N_s}\bold{J}_{\!s}^T\delta\bold{d}_s=-[(\G^T\Delta\bold{d})\circ \S^t]\bold{e}.
\end{align}
In this equation, $\Delta\mathbf{d}$ represents the data residual matrix, and $\mathbf{e}$ is an $N_s$-length column vector consisting of all ones. Each column of $\Delta\mathbf{d}$ corresponds to the data residual associated with a particular source:
\begin{equation}
\Delta\bold{d}=\begin{bmatrix}
\delta\bold{d}_1& \delta\bold{d}_2 & \cdots & \delta\bold{d}_{N_s}
\end{bmatrix}.
\end{equation}
To simplify the $i$th component of the gradient vector, we apply the relation defined in \eref{Hadamard_diag}:
\begin{align}  \label{FWI_grad}
\left[\sum_{s=1}^{N_s}\bold{J}_{\!s}^T\delta\bold{d}_s\right]_i=-[\G^T\Delta\bold{d}\S]_{i},
\end{align}
where $i=1,\ldots,N$. Additionally, the GN Hessian matrix can be rewritten as:
\begin{equation} \label{GN_Hessian}
\sum_{s=1}^{N_s}\bold{J}_{\!s}^T\bold{J}_{\!s}=(\G^T\G) \circ (\S^T\S).
\end{equation} 
By utilizing \eref{Hadamard_diag}, we can rewrite the GN system in \eref{Gauss_Newton} in an equivalent matrix equation form:
\begin{align} \label{Extended_GN_diag}
[\G^T\G\text{diag}(\delta\bold{m}) \S^T\S]_{ii}= [\G^T\Delta\bold{d} \S]_{ii},
\end{align}
for $i=1,\ldots,N$. Solving the standard GN system, \eref{Gauss_Newton}, is indeed equivalent to solving \eref{Extended_GN_diag} for $\delta\mathbf{m}$. 
So far, we have simply reformulated the GN system into an alternative expression. While solving \eref{Extended_GN_diag} remains a challenging task, the newly formulated representation offers increased flexibility, particularly when considering extensions to the model.
The EGN method departs from \eref{Extended_GN_diag} by relaxing the diagonal constraint on the matrix $\text{diag}(\delta\mathbf{m})$ and considering the full equation. We introduce the extended search direction $\Delta\mathbf{m}$, allowing its off-diagonal coefficients to have nonzero values. Therefore, the EGN search direction solves the following Sylvester matrix equation:
\begin{align} \label{Extended_GN}
\G^T\G \Delta\bold{m} \S^T\S= \G^T\Delta\bold{d} \S.
\end{align}
It is worth noting that this equation reveals the separability of the Hessian matrix in the EGN formulation, making it easier to invert compared to the non-separable GN Hessian in the original formulation, \eref{Gauss_Newton}. Also, \eref{Extended_GN} is obtained due to the full relaxation of the diagonal constraint, however, partial relaxations are also possible but may necessitate more intricate equation manipulation \citep[e.g., ][]{Hou_2015_AIE}.  

A regularization term can be introduced to penalize the off-diagonal elements when solving \eref{Extended_GN}. However, it should be noted that this approach can be computationally and memory intensive due to the size of the resulting solution, which is of size $N \times N$. To explicitly solve \eref{Extended_GN}, we obtain:
\begin{align} \label{dm_ext}
\Delta\bold{m}= (\G^T\G + \mu_S \bold{I})^{-1}\G^T\Delta\bold{d} \S(\S^T\S + \mu_U\bold{I})^{-1},
\end{align}
where $\mu_S$ and $\mu_U$ are damping parameters greater than zero. These parameters ensure a unique solution and help mitigate the ill-conditioning of the matrix inversion.
Using the identity in \eref{Guttman_identity_sub}, we can rewrite \eref{dm_ext} as:
\begin{align} \label{Delta_m}
\Delta\bold{m}
&=\G^T \Delta\bold{d}^e \S, 
\end{align}
where $\Delta\mathbf{d}^e$ is defined as:
\begin{align} \label{deblur}
\Delta\bold{d}^e = \bold{H}_r^{-1}\Delta\bold{d}\bold{H}_s^{-1}.
\end{align}
Here, $\mathbf{H}_r$ and $\mathbf{H}_s$ are the receiver-side and source-side Hessian matrices, respectively, given by:
\begin{equation} \label{Hessians}
\bold{H}_r=\G\G^T + \mu_S \bold{I},\quad \bold{H}_s=\S\S^T + \mu_U \bold{I}.
\end{equation}
Derived from \eref{Delta_m}, we observe that the coefficients of $\Delta\mathbf{m}$ can be straightforwardly  calculated by performing zero-lag cross-correlation between the forward-propagated source-side wavefields and the receiver-side wavefields obtained by back-propagating $\Delta\mathbf{d}^e$.
Consequently, the primary focus of EGN lies in the calculation of the extended data residual $\Delta\mathbf{d}^e$ during the inversion
process. This is achieved by deblurring $\Delta\mathbf{d}$ along the source and receiver dimensions using the inverses of the Hessian matrices $\mathbf{H}_s$ and $\mathbf{H}_r$.

\subsection{Interpretation of the extended direction} 
\begin{figure}[!ht]
\center
\includegraphics[width=0.5\columnwidth]{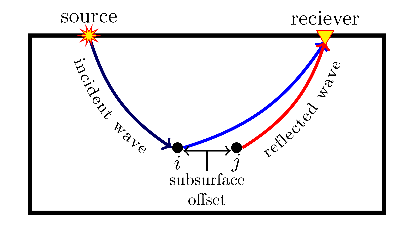}
\caption{Simplified subsurface model with a single source and receiver. The black wave represents the incident wave propagating from the source to the $i$-th reflecting point in the subsurface. At the receiver location, two reflected waves are observed. The blue wave represents the reflection from the same $i$-th point, characterized by the diagonal reflection coefficient $\Delta \bold{m}_{ii}$. The red wave represents a reflection from the $j$-th point, characterized by the off-diagonal reflection coefficient $\Delta \bold{m}_{ij}$. }
\label{imagepoint}
\end{figure}

\noindent The solution $\delta\bold{m}$ obtained from \eref{Extended_GN_diag} also satisfies the linearized (Born) modeling equation in the least squares sense:
\begin{align} \label{Extended_Born1}
\G\text{diag}(\delta\bold{m})\S^T= \Delta\bold{d}.
\end{align}
In this equation, the $(r,i)$-th element of matrix $\G$ represents the Green's function between the $r$-th receiver and the $i$-th subsurface point, and the $(s,j)$-th element of matrix $\S$ represents the Green's function between the $s$-th source and the $j$-th subsurface point. Since only the diagonal elements of $\text{diag}(\delta\bold{m})$ can be non-zero, the scattered data matrix $\Delta\bold{d}$ is constructed only from those source wavefields that arrive at and reflect from the same subsurface point. This is shown in Figure \ref{imagepoint}.

By the same token, the solution $\Delta\bold{m}$ obtained from \eref{Extended_GN} also satisfies the extended Born modeling equation in the least squares sense:
\begin{align} \label{Extended_Born}
\G\Delta\bold{m}\S^T= \Delta\bold{d}.
\end{align}
In this case, the off-diagonal elements of $\Delta\bold{m}$ also contribute to the formation of the data residual matrix. The diagonal elements $\Delta\bold{m}_{ii}$ correspond to the wavefields arriving at and reflecting from the same subsurface point, while the off-diagonal elements $\Delta\bold{m}_{ij}$ (where $i\neq j$) correspond to wavefields arriving at the $i$-th point but reflecting from the $j$-th point. The distance between these subsurface points can be interpreted as the subsurface offset \citep{Claerbout_1985_IEI, Symes_2008_MVA}. Thus, the EGN formulation relaxes the constraint of diagonality and provides an explanation of extended Born modeling from a linear algebra perspective.

\begin{figure}[!ht]
\center
\includegraphics[width=0.5\columnwidth]{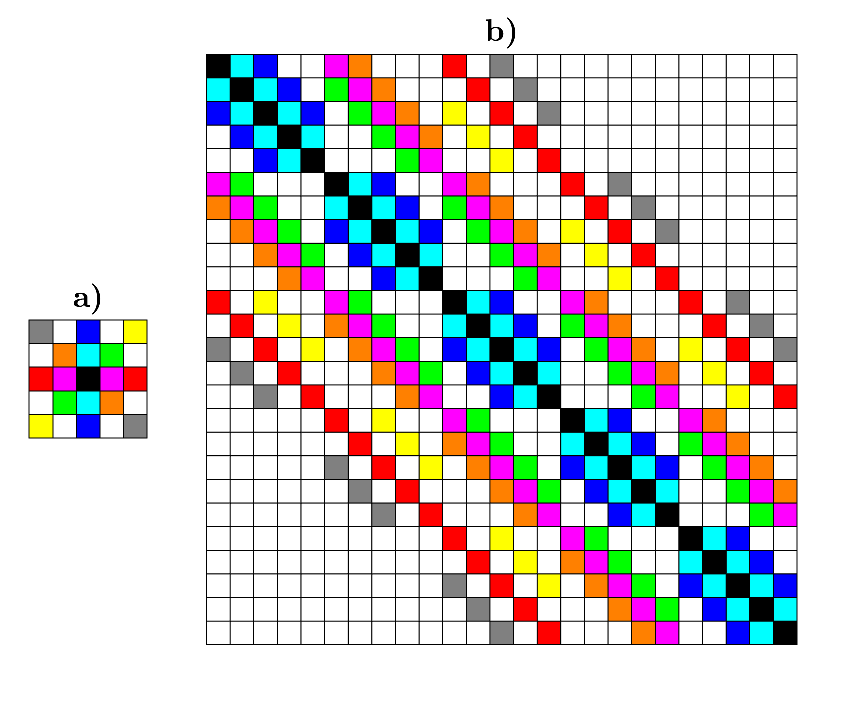}
\caption{Subsurface model and pattern in matrix $\Delta\bold{m}$.
(a) Visualization of a 5 by 5 subsurface model. The black cell (3, 3) serves as the reference cell. Different colors are assigned to cells neighboring (3, 3) along different directions and distances, highlighting their relationships in the model.
(b) The corresponding pattern in matrix $\Delta\bold{m}$, constructed to represent the relationships and positions of cells in the subsurface model. Matrix $\Delta\bold{m}$ has a size of 25 by 25, with each column containing a reshaped column vector from the subsurface model, aligning the respective reference cell along the main diagonal of $\Delta\bold{m}$. The pattern in $\Delta\bold{m}$ reveals the alignment of neighboring cells based on their distances and directions from the reference cell (3, 3).}
\label{matrix}
\end{figure}

Figure \ref{matrix} illustrates the subsurface model and the corresponding pattern in matrix $\Delta\bold{m}$. Here is a clear description of the visualization:
In Figure \ref{matrix}a, we have a 5 by 5 subsurface model. The reference cell, (3, 3), is shown in black. Different colors are assigned to cells that are neighboring (3, 3) along different directions and distances.
The coloring scheme is as follows:
\begin{itemize}
\item Cells (3, 2) and (3, 4) are colored magenta, representing the immediate horizontal neighbors of (3, 3).
\item Cells (3, 1) and (3, 5) have a larger horizontal subsurface offset and are colored red.
\item Cells (2, 3) and (4, 3) are colored cyan, representing the immediate vertical neighbors of (3, 3).
\item Cells (1, 3) and (5, 3) have a larger vertical subsurface offset and are colored blue.
\item  Different colors are also assigned to cells along diagonal directions.
\end{itemize}

In Figure \ref{matrix}b, we visualize matrix $\Delta\bold{m}$, which has a size of 25 by 25 to accommodate the pattern. The goal is to represent the relationships and positions of cells in the subsurface model through the pattern in $\Delta\bold{m}$.
To achieve this, we reshape the original 5 by 5 matrix into a column vector and assign it to column 13 of $\Delta\bold{m}$, ensuring that the reference cell (3, 3) aligns along the main diagonal of $\Delta\bold{m}$. We repeat this procedure for all values of $i$ in the range 1 to 5 and $j$ in the range 1 to 5, placing each reshaped column vector in the associated column of $\Delta\bold{m}$ such that cell $(i, j)$ aligns along the main diagonal.
By following this process, the cells neighboring (3, 3) in the subsurface, which were assigned different colors, will be aligned along different sub-diagonals of matrix $\Delta\bold{m}$. This pattern in $\Delta\bold{m}$ reflects the relationships and positions of cells in the subsurface model, allowing us to visually understand their distances and directions from the reference cell (3, 3). 

Generally, the range and orientation of the subsurface offsets determine how the nonzero patterns in $\Delta\bold{m}$ change. It takes a lot of memory and computational power to solve the associated system for a general pattern. \citet{Hou_2015_AIE} proposed an effective asymptotic (high-frequency) approximate inverse that can work for a limited number of subsurface offsets (partial relaxation). 
 By completely sampling the subsurface offset in range and orientation, we are allowed to fully relax the diagonal constraint in this paper. This is accomplished by allowing nonzero values for all of the coefficients in the matrix $\Delta\bold{m}$. With the diagonal constraint relaxed, the system can be solved more simply and explicitly, which reduces the computational and memory demands of the conventional extended Born equation. This method offers a more thorough and precise solution for the entire range of subsurface offsets.

It should be noted that the emphasis of this paper is primarily on the full relaxation of the diagonal constraint in matrix $\Delta\bold{m}$ which allows for an explicit and exact solution for the associated system. While partial relaxation can certainly be applied by considering a variety of subsurface offsets, it is beyond the scope of this study. 
Further investigation and analysis would be required to determine the challenges associated with partial relaxation, which could be addressed in broader studies or future research specifically focused on exploring the range and relaxation considerations.

\subsection{Computing $\delta \bold{m}$ from $\Delta \bold{m}$}
\noindent 

The proposed algorithm offers a significant advantage in terms of memory efficiency for implementation of model extension. The construction of the extended search direction matrix $\Delta\bold{m}$ is typically challenging due to its large size of $N\times N$. However, if we have access to $\Delta\bold{d}^e$, each individual coefficient of $\Delta\bold{m}$ can be computed independently by cross-correlating the corresponding forward-propagated source-side wavefield with the backward-propagated receiver-side wavefield. 
Thus, compared to the original GN, EGN will not significantly increase the memory requirement. This property
is a key advantage of our algorithm compared to existing methods that rely on model extension, which require storing the model for the entire desired range of subsurface offsets in memory \citep{Rickett_2002_OAC,Biondi_2002_PIO,Biondi_2004_ADC,Symes_2008_MVA,Fu_2017_AMA,Yang_2019_LSR}. 

We compute the desired search direction $\delta\bold{m}$ by averaging the extended coefficients over a ball surrounding each physical point. This approach is based on the observation that the reflection energy tends to interfere constructively within the first Fresnel zone \citep{Sheriff_1980_NFC,Lindsey_1989_FZI}.
It is worth noting that a similar idea was employed by \citet{Yang_2019_LSR}, where they averaged the coefficients over a horizontal disk centered at the physical point. Our approach extends this concept by considering a ball instead of a disk, enabling more comprehensive averaging and capturing the interference effects in all directions around each physical point.

The computation of the search direction is described by 
\begin{equation} \label{model_update}
\delta\bold{m}(\bold{x})= \text{Re}\left(\sum_{s} \sum_{\omega}\sum_{\bold{h}} \phi(\bold{h})\left(\frac{\partial \bold{A(m)}}{\partial \bold{m}}\bold{u}(\omega,\bold{x}- \bold{h};\bold{x}_s)\right)^T
\bold{v}(\omega,\bold{x}+\bold{h};\bold{x}_s)\right).
\end{equation}
Assuming that the extended data residual $\Delta\bold{d}^e$ in equation \ref{deblur} has been correctly computed, the search direction can be obtained by performing a multiplication operation between the source and receiver wavefields.
We can see that the computed wavefields can be space-shifted to compute the extended coefficient corresponding to each subsurface half-offset $\bold{h}$, eliminating the need to solve additional simulation problems for each background model.

In detail, the equation involves the following steps:
\begin{enumerate}
\item The source wavefield $\bold{u}(\omega,\bold{x};\bold{x}_s)$ is computed for each source location $\bold{x}_s$ and for a given frequency $\omega$. This wavefield represents the wave propagation from the physical source $\bold{b}(\omega,\bold{x};\bold{x}_s)$ to the subsurface location $\bold{x}$. They are multiplied by the radiation pattern matrix $\frac{\partial \bold{A(m)}}{\partial \bold{m}}$. We keep this form 
for different parameterizations.

\item The adjoint source $\bold{P}^T\delta\bold{d}^e(\omega;\bold{x}_s)$ is used to generate the receiver wavefield $\bold{v}(\omega,\bold{x};\bold{x}_s)$. This adjoint wavefield represents the wave propagation from the receiver locations to the subsurface location $\bold{x}$.

\item The product of these wavefields is computed and summed over the desired subsurface half-offset vector $\bold{h}$, frequency range, and source range. $\phi(\bold{h})$ is a weight function \citep{Hou_2015_AIE}.
It should be noted that the effect of the least squares solve of the extended system is encoded directly in the extended data residual, $\Delta\bold{d}^e$. As a result, there is no need to calculate and store individual subsurface offset coefficients and perform separate averaging operations. Instead, the averaging process is performed on-the-fly as soon as the two wavefields are available.

\item The real part of the resulting quantity is taken, representing the search direction at the subsurface location $\bold{x}$.
\end{enumerate}

\subsection{Step length selection}
\noindent After computing the step direction $\delta\bold{m}$, the next step is to determine the step length $\alpha$ for updating the model. The ideal choice for $\alpha$ would be the global minimizer of the data misfit function along the step direction. However, this minimizer is generally difficult to obtain, and approximate minimizers are typically used in practice.
One approach to approximate the minimizer is through standard line search methods, which involve iteratively testing different step lengths and selecting the one that yields the lowest misfit.
Alternatively, an approximate minimizer can be obtained by setting the gradient of the misfit function with respect to $\alpha$ equal to zero, which transforms the step length determination into a fixed-point problem. This is done using the equation:
\begin{equation} \label{fixedpoint}
\alpha=\frac{\langle \G(\alpha)\text{diag}(\delta\bold{m})\S(\alpha)^T, \Delta\bold{d}\rangle_F}{\langle\G(\alpha)\text{diag}(\delta\bold{m})\S(\alpha)^T, \G(0)\text{diag}(\delta\bold{m})\S(\alpha)^T\rangle_F}=:\varphi(\alpha),
\end{equation}
where $\G(\alpha)$ and $\S(\alpha)$ are constructed by $\bold{m}_k+\alpha \delta\bold{m}$. This equation defines a function $\varphi(\alpha)$ that represents the fixed-point iteration for finding a local minimizer of the misfit function.

The fixed-point iteration is performed by starting with an initial guess $\alpha_0 = 0$ and iteratively updating $\alpha$ using the equation:
\begin{equation} \label{FPT}
\alpha_{l+1} = \varphi(\alpha_l),
\end{equation}
where $l$ represents the iteration number. The first iteration, $\alpha_1$, corresponds to the step length proposed by \citet{Pratt_1998_GNF}. In cases where the nonlinearity of the data is not significant, the first iteration often provides an accurate estimate of the step length. However, when the data nonlinearity is high, more iterations may be required to improve the estimate of the step length.

\subsection{Randomized EGN}
\noindent In order to reduce the computational cost associated with solving the EGN system, \eref{Extended_Born}, in a least squares sense, we employ random projection schemes \citep{Wang_2017_SMR}. These schemes allow us to effectively reduce the dimensions of the system, making it more computationally efficient to solve.

We introduce a random projection (sketching) matrix $\bold{\Pi}_{a\times b}$ of size $a\times b$, where $b$ is much smaller than $a$, and it satisfies the property 
\begin{equation}
\mathbb{E}(\bold{\Pi}_{a\times b}\bold{\Pi}_{a\times b}^T)=\bold{I},
\end{equation}
where $\mathbb{E}$ denotes the expectation. By applying this projection matrix, we can work with the projected matrices $\G_{\Pi}$ and $\S_{\Pi}$, obtained by multiplying the original matrices $\G$ and $\S$ by $\bold{\Pi}^T_{N_r\times N_p}$ and $\bold{\Pi}^T_{N_s\times N_q}$.  
This compresses the data residual matrix, $\Delta\bold{d}$, of size $N_r\times N_s$ to $\Delta\bold{d}_{\Pi} =\bold{\Pi}^T_{N_r\times N_p}\Delta\bold{d}\bold{\Pi}_{N_s\times N_q}$ of size $N_p\times N_q$.

Various sketching matrices can be used for this purpose, such as random Gaussian sketch, random phase sketch, random sampling, and count sketch \citep[see][ for a complete review]{Aghazade_2021_RSS}. The main advantage of using random projections is that the resulting sketched system, given by \eref{Extended_GNr}, has a significantly smaller dimension compared to the original system. This reduction is particularly beneficial for the receiver dimension, as the number of receivers is typically much larger than the number of sources in practical settings. By leveraging random projections, we can achieve computational efficiency by effectively reducing the size of the system and solving the sketched version of the EGN system, i.e.
\begin{align} \label{Extended_GNr}
\G_{\Pi} \Delta\bold{m} \S_{\Pi}^T= \Delta\bold{d}_{\Pi}.
\end{align}

\subsection{Computational complexity}
\noindent When comparing with the computational complexity of the steepest-descent method, the primary computational cost in our approach lies in the deblurring step of the data residual matrix, as described in equations \ref{deblur} or \ref{deblur_penalty}. This step involves the computation of two Hessian matrices, $\bold{H}_s$ and $\bold{H}_r$, and the application of their inverses to the source and receiver coordinates of the data residual matrix.
The size of these matrices is $N_s \times N_s$ for $\bold{H}_s$ and $N_r \times N_r$ for $\bold{H}_r$. Fortunately, these matrices can be constructed explicitly and inverted using direct methods. The construction of $\bold{H}_s$ does not impose additional computational cost since the required forward wave-equation solves are already available from computing the data residuals.
To build $\bold{H}_r$, we only need $N_r$ additional backward wave-equation solves. However, by employing the random sketching strategy discussed earlier, we can decrease the total number of required wave-equation solves. In this case, we only need $N_p$ extra wave-equation solves compared to the number required for the steepest-descent method.
By utilizing the random sketching approach, we achieve a significant reduction in computational burden by minimizing the number of wave-equation solves required for the deblurring step. This reduction in computational cost enhances the efficiency and practicality of our approach while still preserving its extension properties for robust inversion.

\section{Application of EGN to extended source FWI}
\noindent In the extended source FWI framework, we aim to increase the robustness of waveform inversion by incorporating wave-equation relaxation. While the objective function associated with this approach can be formulated in bivariate form \citep{Abubakar_2009_FDC,VanLeeuwen_2013_MLM,Huang_2018_VSE}, it can also be defined in a univariate form in the reduced space by employing variable projection. This results in a penalty objective function expressed as a weighted misfit norm via
\begin{equation} \label{primal_penalty_red}
E_{\beta}(\bold{m})=\frac12\sum_{s=1}^{N_s}\|\bold{F}_{\!\!s}(\bold{m})-\bold{d}_s\|_{\bold{Q}(\bold{m})^{-1}}^2,
\end{equation}
\citep{vanLeeuwen_2019_ANO,Gholami_2022_EFW}.
The penalty objective function, denoted as $E_{\beta}(\bold{m})$, is defined as the sum of weighted misfit norms of the data residuals. The model-dependent weights are determined by the inverse of the covariance matrix $\bold{Q}(\bold{m})$. Specifically, $\|\bold{y}\|_{\bold{Q}(\bold{m})}^2=\bold{y}^T\bold{Q(\bold{m})}\bold{y}$ denotes the weighted norm of vector $\bold{y}$ with respect to $\bold{Q}(\bold{m})$, and $\bold{Q}(\bold{m})$ is defined as 
\begin{equation}\label{Q}
\bold{Q}(\bold{m})=\frac{1}{\beta}\G(\bold{m})\G(\bold{m})^T + \bold{I},
\end{equation}
where $\beta$ is a penalty parameter that penalizes wave-equation relaxation. It plays a crucial role in controlling the amount of wave-equation relaxation in the inversion process. By increasing the value of $\beta$, we impose a stronger penalty on the wave-equation constraint, effectively reducing the relaxation and approaching the original waveform inversion problem. At the beginning of the inversion process, a small value is assigned to $\beta$. As the iterations progress, the value of $\beta$ may be increased gradually, following a penalty continuation strategy. In the limit as $\beta$ approaches infinity, the penalty objective function $E_{\beta}$ reduces to the original objective function $E$, which represents the standard waveform inversion without wave-equation relaxation.

To minimize $E_{\beta}(\bold{m})$, standard gradient-based optimization methods can be employed. The GN method leads to a linear system for the search direction $\delta\bold{m}$ as \citep{vanLeeuwen_2016_PMP,Gholami_2022_EFW}
\begin{equation} \label{Gauss_Newton_WRI}
\left[\sum_{s=1}^{N_s}\bold{J}_{\!s}^T\bold{Q}^{-1}\bold{J}_{\!s}\right] \delta\bold{m}=-\sum_{s=1}^{N_s}\bold{J}_{\!s}^T\bold{Q}^{-1}\delta\bold{d}_s.
\end{equation}
In this equation, the Jacobian matrix $\bold{J}_{\!s}$ is defined similar to \eref{J}, but with the extended wavefields $\bold{u}^{\beta}_s$, which are defined as
\begin{subequations}\label{extended}
\begin{align} 
\bold{u}^{\beta}_s&=\arg\min_{\bold{u}} ~\|\bold{Pu}-\bold{d}_s\|_2^2 + \beta\|\bold{A}\bold{u}-\bold{b}_s\|_2^2 \label{extended_u}\\
& =\bold{A}^{-1}(\bold{b}_s-\frac{1}{\beta}\G^T\bold{Q}^{-1}\delta\bold{d}_s). \label{ue_2}
\end{align}
\end{subequations}
By transforming \eref{Gauss_Newton_WRI} into an equivalent matrix equation form, we arrive at the following equation which can be solved for the search direction: 
\begin{align} \label{Extended_GN_diag_pemalty}
[\G^T\bold{Q}^{-1}\G\text{diag}(\delta\bold{m}) \S^T_{\beta}\S_{\beta}]_{ii}= [\G^T\bold{Q}^{-1}\Delta\bold{d} \S_{\beta}]_{ii},
\end{align}
for $i=1,...N$.
In this equation, the matrix $\S_{\beta}$ includes the extended wavefields as its rows and is explicitly defined as
\begin{equation}\label{U_Lambda}
\bold{U}_{\beta}= \omega^2\left[ \bold{B}-\frac{1}{\beta}\G^T\bold{Q}^{-1}\Delta\bold{d}\right]^T\bold{A}^{-T}.
\end{equation}
The EGN search direction for the extended source FWI is obtained by solving
\begin{equation} \label{Gauss_Newton_WRI_ext}
\G^T\bold{Q}^{-1} \G\Delta\bold{m} \S^T_{\beta}\S_{\beta} = \G^T \bold{Q}^{-1}\Delta\bold{d} \S_{\beta}.
\end{equation}
By using the matrix equalities in equations \ref{Guttman_identity} and \ref{Guttman_identity_sub} and the definition of $\bold{Q}$, equation \ref{Q}, we get the following explicit solution for equation \ref{Gauss_Newton_WRI_ext}:
\begin{equation} \label{Delta_m_redp}
\Delta\bold{m} = \G^T \Delta\bold{d}^e\S_{\beta},
\end{equation}
where
\begin{equation} \label{deblur_penalty}
\Delta\bold{d}^e = \bold{H}_r^{-1}\Delta\bold{d}\bold{H}_s^{-1}.
\end{equation}
The Hessian matrices $\bold{H}_r$ and $\bold{H}_s$ are defined as
\begin{equation} \label{Hrs_penalty}
\bold{H}_r=\epsilon\left(\G\G^T + \epsilon\mu_S\bold{I}\right),\quad \bold{H}_s=\S_{\beta}\S_{\beta}^T + \mu_U \bold{I},
\end{equation}
where $\epsilon=\beta/(\beta+\mu_S)$.
As expected, equations \ref{Delta_m_redp}-\ref{Hrs_penalty} reduce to equations \ref{Delta_m}-\ref{Hessians} for $\beta\to \infty$.

In the context of waveform reconstruction inversion (WRI), a sparse approximation is commonly employed for the GN Hessian matrix $\bold{J}_{\!s}^T\bold{Q}^{-1}\bold{J}_{\!s}$  \citep{vanLeeuwen_2016_PMP}. This approximation replaces the full Hessian matrix in \eref{Gauss_Newton_WRI} with a sparse approximation, given by $\beta\text{diag}(\omega^2 \bold{u}_s^{\beta})^T\text{diag}(\omega^2 \bold{u}_s^{\beta})$. When applying this sparse approximation, we observe that the extended search direction in WRI remains defined by equations \ref{Delta_m_redp} to \ref{Hrs_penalty}. However, there is a simple modification in the receiver-side Hessian matrix $\bold{H}_r$, which becomes $\bold{H}_r = \G\G^T + \beta\bold{I}$.

For the ease of comparison, the extended search direction and associated parameters for both the reduced FWI and penalty FWI are summarized in Table \ref{Table}.
 The algorithms for these two cases are very similar, with the only difference lying in their incident wavefields, which in turn lead to different source-side wavefield matrices: $\S$ for standard FWI and $\S_{\beta}$ for penalty FWI. This distinction arises from the concept of extended sources, as defined in \eref{U_Lambda}: 
 $\bold{b}_s-\frac{1}{\beta}\G^T\bold{Q}^{-1}\delta\bold{d}_s$.
The extended source consists of two terms. The first term represents the primary source, $\bold{b}_s$, while the second term represents a secondary source, denoted as $\boldsymbol{\varphi}_s=-\frac{1}{\beta}\G^T\bold{Q}^{-1}\delta\bold{d}_s$. These secondary sources are obtained by solving the scattered data equation in a least squares sense: $\G\boldsymbol{\varphi}_s=-\delta\bold{d}_s$.
The interpretation of this equation is straightforward: the data residuals from the background model are inverted in a least squares sense to obtain the corresponding scattering source. The damping parameter $\beta$ plays a crucial role in this process, allowing us to find a minimum energy source that explains the scattered data. When this estimated scattering source is added to the physical source, it partially accounts for multiscattering effects and leads to an improved linearization of the FWI forward problem \citep{Gholami_2022_OCN}. However, this improvement in the incident wavefield is obtained by additional wave-equation solves compared to the reduced wavefield used in reduced FWI.
The choice of a large damping parameter effectively removes the contribution of these secondary sources, reducing the extended wavefields to the reduced wavefields, which correspond to the standard FWI case.

\begin{table*}[ht!]
\begin{center}
\caption{The formula of extended search direction and the associated parameters for the EGN method applied to the FWI problem with reduced objective, equation \ref{main_obj}, and  penalty objective, equation \ref{primal_penalty_red}.}\label{Table}
\vspace{.3cm}
\small 
\renewcommand{\arraystretch}{1.2}
\begin{tabular}{|c|c|c|c|c|c|c|}
\hline
 & \multicolumn{6}{c|}{$\Delta\bold{m} = \epsilon\G^T(\G\G^T + \epsilon\mu_S \bold{I})^{-1}\Delta\bold{d}  (\S\S^T + \mu_U\bold{I})^{-1} \S $} \\
 \hline \hline
 \cline{2-7} \cline{2-7} 
method& $\epsilon$ & $\G$ & $\S_{s:}$ & $\bold{u}_s$ & $\delta\bold{b}_s$ & $\delta\bold{d}_s$  \\
\hline 
reduced  & 1 & $\bold{PA}^{-1}$ & $\omega^2 \bold{u}^T_s$ & $\bold{A}^{-1}[\bold{b}_s+\delta\bold{b}_s]$ & $\bold{0}$ & $\G\bold{b}_s-\bold{d}_s$ \\
\hline
penalty & $\frac{\beta}{\beta+\mu_S}$ & $\bold{PA}^{-1}$ & $\omega^2 \bold{u}^T_s$ & $\bold{A}^{-1}[\bold{b}_s+\delta\bold{b}_s]$ & $-\G^T(\G\G^T + \beta \bold{I})^{-1}\delta\bold{d}_s$ & $\G\bold{b}_s-\bold{d}_s$ \\
\hline
\end{tabular}
\normalsize
\end{center}
\end{table*}

\section{NUMERICAL EXAMPLES}
\noindent In this section, we assess the performance of the EGN method on various 2D monoparameter synthetic examples. We consider different acquisition settings, including transmission data inversion, reflection-dominated data from surface acquisition, simultaneous inversion of multiple frequencies, and multiscale inversion using frequency continuation.

To enable simultaneous inversion of multiple frequencies, we need to solve a matrix equation known as the generalized Sylvester equation \citep{Hautus_1994_OS}. This equation takes the form:
\begin{align} \label{GSE}
\sum_{\omega}\G_{\omega}^T\G_{\omega} \Delta\bold{m} \S_{\omega}^T\S_{\omega}= \sum_{\omega}\G_{\omega}^T\Delta\bold{d}_{\omega} \S_{\omega},
\end{align}
where the subscript ${\omega}$ indicates the frequency dependence. It is important to note that the range of $\omega$ should be symmetric to ensure a real-valued result and to maintain consistency with the fact that the gradient of the real-valued misfit function with respect to real-valued parameters is also real. However, solving this generalized Sylvester equation efficiently, even for two frequencies, is challenging and remains an open research problem. In this study, we address this issue by separately and in parallel inverting different positive frequencies. The EGN step direction is then obtained by averaging the results from individual frequencies and taking its real value.

To emphasize the effect of the search direction extension, we refrain from applying specific regularization or smoothing to the step direction. Additionally, in the EGN algorithm, the damping parameters $\mu_S$ and $\mu_U$ in the Hessian matrices are set to be a small fraction (0.01) of the maximum eigenvalue of their respective matrices. The step length is determined automatically at each iteration using the fixed-point iteration method described by \eref{FPT}. For the tests conducted in this study, we use the step length provided by the first iteration.
This makes the inversion free of parameters for fine-tuning and automates the selection of the essential EGN parameters. By automating the parameter selection procedure, the algorithm strikes a reasonable balance between efficiency and robustness. It decreases the dependence on user expertise while increasing the consistency and effectiveness of the inversion process.

In our numerical examples, we employ the following inversion methods:
\begin{enumerate}
\item PSD method: The search direction is determined by \eref{Gauss_Newton} while approximating the GN Hessian with a sparse approximation, known as the pseudo-Hessian matrix, $\mathbf{J}_{\!s}^T\mathbf{J}_{\!s}\approx\text{diag}(\omega^2 \bold{u}_s)^T\text{diag}(\omega^2 \bold{u}_s)$ \citep{Shin_2001_IAP}. This approximation only considers the diagonal elements of the source-side Hessian matrix $\S^T\S$ and neglects the diagonal elements of the receiver-side Hessian matrix $\G^T\G$. Thus, it partially accounts for the geometrical spreading phenomenon.

\item GN method: The search direction is determined by the GN formulation in \eref{Gauss_Newton}. More specifically, we utilize the damped Hessian $\mathbf{J}_{\!s}^T\mathbf{J}_{\!s} + \mu \bold{I}$, where $\mu$ is equal to 0.01 of the maximum eigenvalue of $\mathbf{J}_{\!s}^T\mathbf{J}_{\!s}$.

\item EGN solving the reduced FWI: The search direction is determined by \eref{model_update}, where the source wavefields are defined as $\bold{u}_s=\bold{A(m)}^{-1}\bold{b}_s$, and the extended data residual $\Delta\mathbf{d}^e$ defining the receiver wavefields is computed using \eref{deblur}.

\item EGN solving the penalty FWI: The search direction is determined by \eref{model_update}, where the source wavefields are defined as in \eref{ue_2} and the extended data residual $\Delta\bold{d}^e$ defining the receiver wavefields is computed using \eref{deblur_penalty}.
\end{enumerate}

We want to emphasize that our main focus in this paper is on achieving complete relaxation of the diagonality constraint and obtaining an exact solution for the resulting EGN system. In most cases, the search direction extracted at zero subsurface offset performs satisfactorily.
 Consequently, in many of our numerical examples, the EGN algorithm primarily utilizes the diagonal coefficients of the extended search direction.  This means that \eref{model_update} is evaluated only for $\bold{h}=\bold{0}$ or using the delta function as the weighting function $\phi(\bold{h})$. 
Note that the EGN method with $\bold{h}=\bold{0}$ differs from the original GN method. 
In the former, the extended search direction $\Delta\bold{m}$ is initially computed by fitting data residuals in a least-squares sense, followed by the extraction of its diagonal as $\delta\bold{m}$. However, in the latter, $\delta\bold{m}$ is directly computed by fitting data residuals in a least-squares sense.
To further support our theoretical framework, we also demonstrate the potential improvement in results by averaging the coefficients within the half-wavelength width. Specifically, we consider the case for $|\bold{h}|\leq \frac{1}{4}\lambda$, where $\lambda$ is the dominant wavelength. In this case, we introduce a simple weight function $\phi(\bold{h})$ that decreases exponentially with $|\bold{h}|$. However, determination of the optimum subsurface offset range and weight function is beyond the scope of this study and is left for future investigation.

\subsection{Camembert model example}
\begin{figure}[!ht]
\center
\includegraphics[width=1\columnwidth]{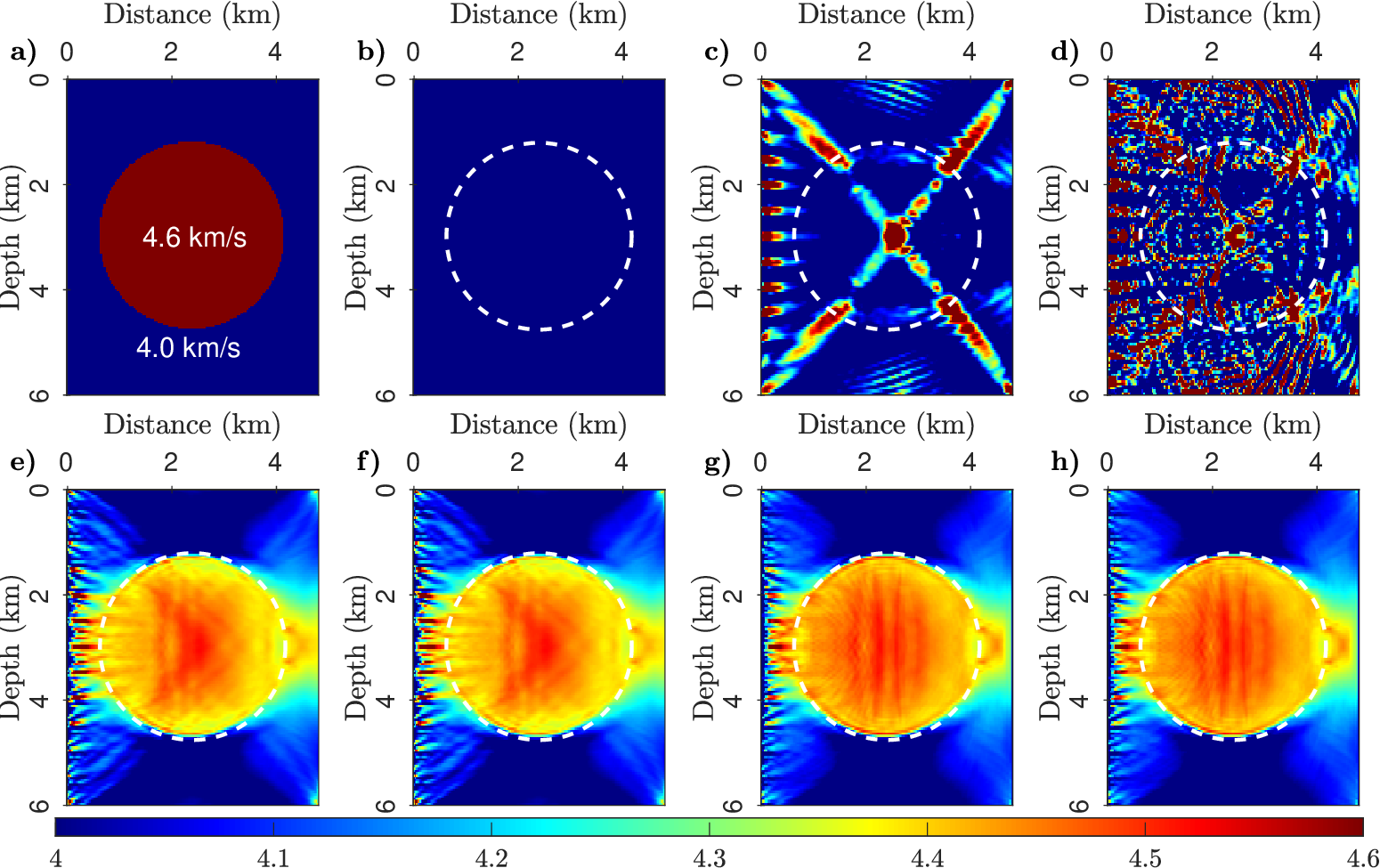}
\caption{Camembert model example. The (a) true velocity model and (b) initial model. (c-h) Final velocity models inferred after 50 iterations of (c) PSD, (d) GN, (e) EGN solving the reduced FWI with $\bold{h}=\bold{0}$, (f) EGN solving the penalty FWI with $\bold{h}=\bold{0}$, (g) EGN solving the reduced FWI with $|\bold{h}|\leq \frac{1}{4}\lambda$, (h) EGN solving the penalty FWI with $|\bold{h}|\leq \frac{1}{4}\lambda$.}
\label{Camembert}
\end{figure}

\noindent We evaluate the performance of the proposed method using the challenging ``Camembert" model \citep{Gauthier_1986_TDN}. This model consists of a circular anomaly with a velocity of 4.6 km/s embedded in a homogeneous background with a velocity of 4.0 km/s (Figure \ref{Camembert}a). The dimensions of the model are 4.8 km in distance and 6 km in depth, with a grid spacing of 35.5 m.
For the crosshole acquisition setup, we deploy 13 equally spaced sources on the left side of the model and 170 equally spaced receivers vertically on the opposite side, simulating a crosshole seismic experiment. The source signature used in this study is a 10 Hz Ricker wavelet.
The presence of a relatively large anomaly diameter makes the inverse problem nonlinear and challenging. The convergence of local optimization algorithms in such cases strongly depends on the initial model. In this study, we initiate the inversion from the homogeneous background model (Figure \ref{Camembert}b) to ensure that the problem remains in the nonlinear regime of FWI. We perform one cycle of inversion, meaning that we do not follow a multiscale approach, and all frequencies in the data (ranging from 3 Hz to 25 Hz with a 1 Hz frequency interval) are inverted simultaneously.

\begin{figure}[!ht]
\center
\includegraphics[width=0.6\columnwidth]{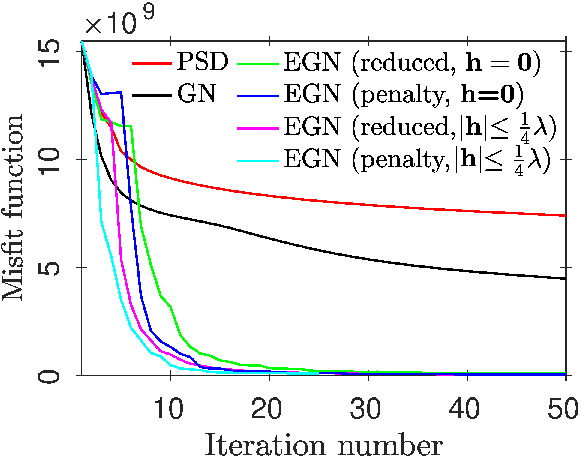}
\caption{Trajectory of the misfit function versus iteration for the Camembert model obtained by different methods (Figures \ref{Camembert}c-f). The curves represent the misfit function evolution for the following methods: PSD (red line), GN (black line), EGN solving the reduced FWI with $\bold{h}= \bold{0}$ (green line), EGN solving the penalty FWI with $\bold{h}= \bold{0}$ (blue line), EGN solving the reduced FWI with $|\bold{h}|\leq \frac{1}{4}\lambda$ (magenta line), and EGN solving the penalty FWI with $|\bold{h}|\leq \frac{1}{4}\lambda$ (cyan line).}
\label{Camembert_conv}
\end{figure}

We compare the velocity models obtained after 50 iterations using different inversion methods: PSD (Figure \ref{Camembert}c), GN (Figure \ref{Camembert}d), EGN solving the reduced FWI (Figure \ref{Camembert}e), and EGN solving the penalty FWI (Figure \ref{Camembert}f), both with $\bold{h}= \bold{0}$. We observe that the classical methods (PSD and GN) fail to recover the model due to severe cycle-skipping. In contrast, the EGN methods effectively mitigate cycle-skipping, resulting in reconstructed models (Figures \ref{Camembert}e-f) that closely resemble the ground truth model. It is worth noting that Figures \ref{Camembert}e and \ref{Camembert}f are visually indistinguishable, but the penalty FWI shows a faster convergence rate, as indicated by the associated misfit function in Figure \ref{Camembert_conv} (blue curve versus green curve). We can also observe from Figure \ref{Camembert_conv} that GN exhibits a faster convergence rate than PSD, but the estimated model (Figure \ref{Camembert}d) contains more artifacts while better fitting the data. This suggests that both PSD and GN converge to a stationary point, which is an undesirable local minimum. Therefore, we do not present the results obtained from the GN method in the subsequent analysis due to its inefficiency.

To demonstrate the effect of using the off-diagonal coefficients of $\Delta\bold{m}$, we run the EGN method while computing $\delta\bold{m}$ by averaging the extended coefficients within the half-wavelength width. The results obtained using the EGN method for both the reduced and penalty FWI with $|\bold{h}|\leq \frac{1}{4}\lambda$ are shown in Figures \ref{Camembert}g-h. By comparing these figures with Figures \ref{Camembert}e-f, we observe that averaging over nonzero subsurface offsets leads to an improvement in the velocity model. This improvement is also evident from the associated convergence curves in Figure \ref{Camembert_conv} (magenta and cyan curves versus blue and green curves).

\begin{figure}[!ht]
\center
\includegraphics[width=1\columnwidth]{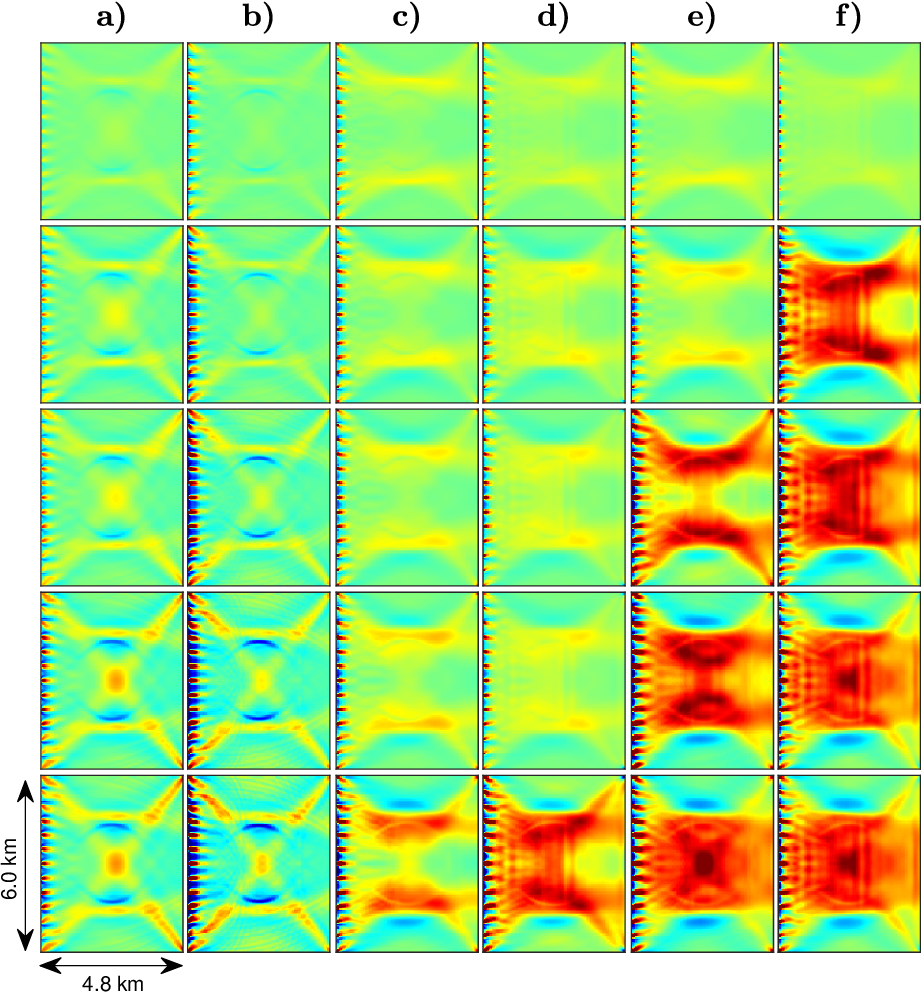}
\caption{Comparison of model perturbation accumulated in the first 5 iterations for different FWI methods.
The figure displays subplots (a) to (f), each corresponding to a different method used in constructing the Camembert model. The rows represent the model perturbation obtained after a specific number of iterations (from the top row, representing the first iteration, to the bottom row, representing the fifth iteration). The methods include (a) PSD, (b) GN, (c) EGN applied to the reduced FWI with $\bold{h}= \bold{0}$, (d) EGN applied to the penalty FWI with $\bold{h}= \bold{0}$, (e) EGN applied to the reduced FWI with $|\bold{h}|\leq \frac{1}{4}\lambda$, and (f) EGN applied to the penalty FWI with $|\bold{h}|\leq \frac{1}{4}\lambda$.
 }
\label{Camembert_dm}
\end{figure}

In Figure \ref{Camembert_dm}, we compare the model perturbation obtained during the first five iterations for various methods used in building the Camembert model. Starting at the top row (first iteration) and moving down (up to the fifth iteration), each row in the subplots represents the model perturbation that was obtained after a specific number of iterations. We can see that the perturbation produced by PSD (Figure \ref{Camembert_dm}a) and GN (Figure \ref{Camembert_dm}b) primarily includes the top and bottom boundaries of the circular anomaly, demonstrating that they become stuck in a local minimum. This observation is in line with the characteristics of these techniques, which rely on the first-order Born scattering theory. 

The EGN method significantly enhances the estimate by gradually building the anomaly from its edge into its center, while still only using the diagonal coefficients of the extended search direction (Figure \ref{Camembert_dm}c). The performance is further enhanced when applied to the penalty FWI due to the improvement in the incident wavefield quality (Figure \ref{Camembert_dm}d).
When the reduced FWI is still used and the off-diagonal coefficients are added, a significant improvement is made (Figure \ref{Camembert_dm}e). When the off-diagonal coefficients are used in the context of EGN solving the penalty FWI, the greatest improvement is seen (Figure \ref{Camembert_dm}f). In this case, a significant portion of the anomaly is recovered even after the first iteration. The ability of EGN to invert multiscattered data as a result of the extension can be used to justify the improvements.

\subsection{Marmousi model example}
\begin{figure}[!ht]
\center
\includegraphics[width=1\columnwidth]{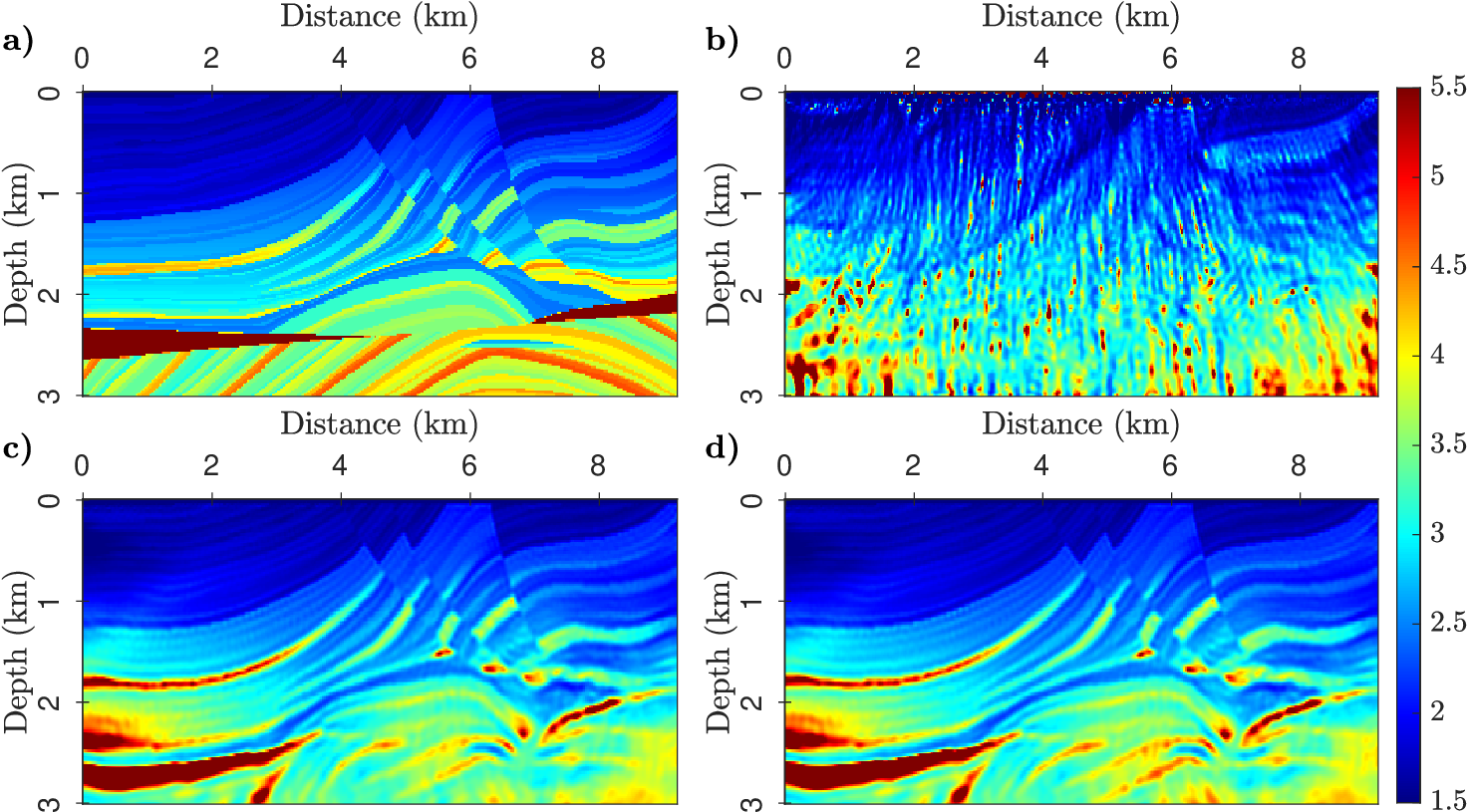}
\caption{Marmousi model example. The (a) true velocity model. Final velocity models inferred by (b) the PSD method, (c) the EGN method solving the reduced FWI with $\bold{h=0}$, and (d) the EGN method solving the penalty FWI with $\bold{h=0}$.}
\label{Marmousi_I}
\end{figure}

\noindent We further evaluate the performance of the proposed EGN method in the inversion of reflection-dominated data using surface acquisition. The true velocity model is shown in Figure \ref{Marmousi_I}a, representing a subsurface model with dimensions of 9.2 km width and 3 km depth, discretized with a 20 m grid interval. The fixed spread acquisition setup consists of 47 equally-spaced sources and 154 equally-spaced receivers located on the top side of the model. To mitigate boundary effects, a perfectly matched layer (PML) absorbing boundary condition is implemented, and a Ricker wavelet with an 8 Hz dominant frequency is used as the source signature.

\begin{figure}[!ht]
\center
\includegraphics[width=0.6\columnwidth]{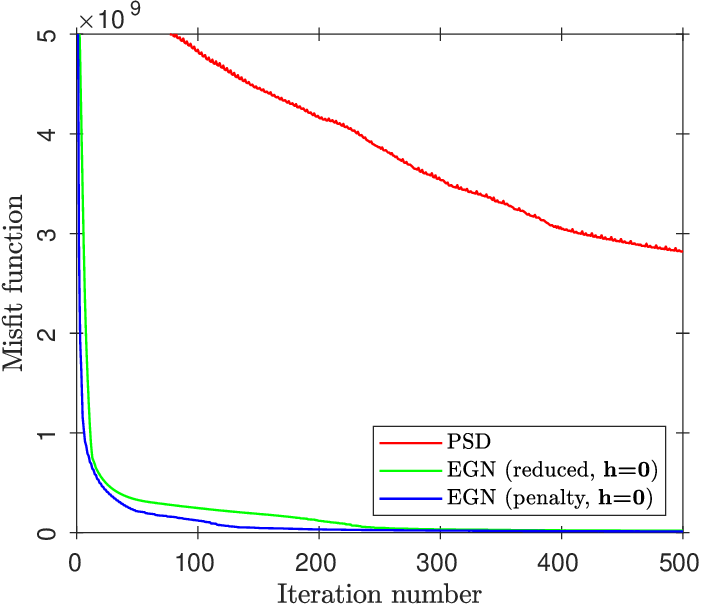}
\caption{Trajectory of the misfit function versus iteration for the Marmousi model obtained by the PSD method (red line), EGN method solving the reduced FWI (green line), and EGN method solving the penalty FWI (blue line).}
\label{MarmousiI_conv}
\end{figure}

We initiate the inversion process with a 1D initial model that linearly increases in depth from 1.5 km/s to 4.0 km/s. Simultaneous inversion of all frequencies in the data is performed, covering the range of 3 Hz to 13 Hz with a frequency interval of 0.5 Hz. The obtained velocity models after 500 iterations of the PSD method and the EGN method, solving the reduced FWI, are presented in Figures \ref{Marmousi_I}b and \ref{Marmousi_I}c, respectively. Additionally, the velocity model resulting from the EGN method solving the penalty FWI is shown in Figure \ref{Marmousi_I}d. The associated convergence behavior of the misfit functions versus iteration is illustrated in Figure \ref{MarmousiI_conv}.

It is evident that the PSD method failed to converge to a reasonable solution, as shown in Figure \ref{Marmousi_I}b. The corresponding misfit curve exhibited slow convergence towards an undesired local minimum. In contrast, both EGN algorithms, starting from the same initial model, successfully converged to accurate solutions, as shown in Figures \ref{Marmousi_I}c and \ref{Marmousi_I}d, with rapid convergence of the associated misfit curves. Notably, the EGN method solving the penalty FWI exhibited even faster convergence compared to the EGN method solving the reduced FWI.
Although not presented here, it is worth mentioning that the results obtained for $|\bold{h}|\leq \frac{1}{4}\lambda$ were similar to those achieved with $\bold{h}= \bold{0}$, as shown in Figures \ref{Marmousi_I}c and \ref{Marmousi_I}d.

\subsection{Overthrust model example}
\begin{figure}[!ht]
\center
\includegraphics[width=1\columnwidth]{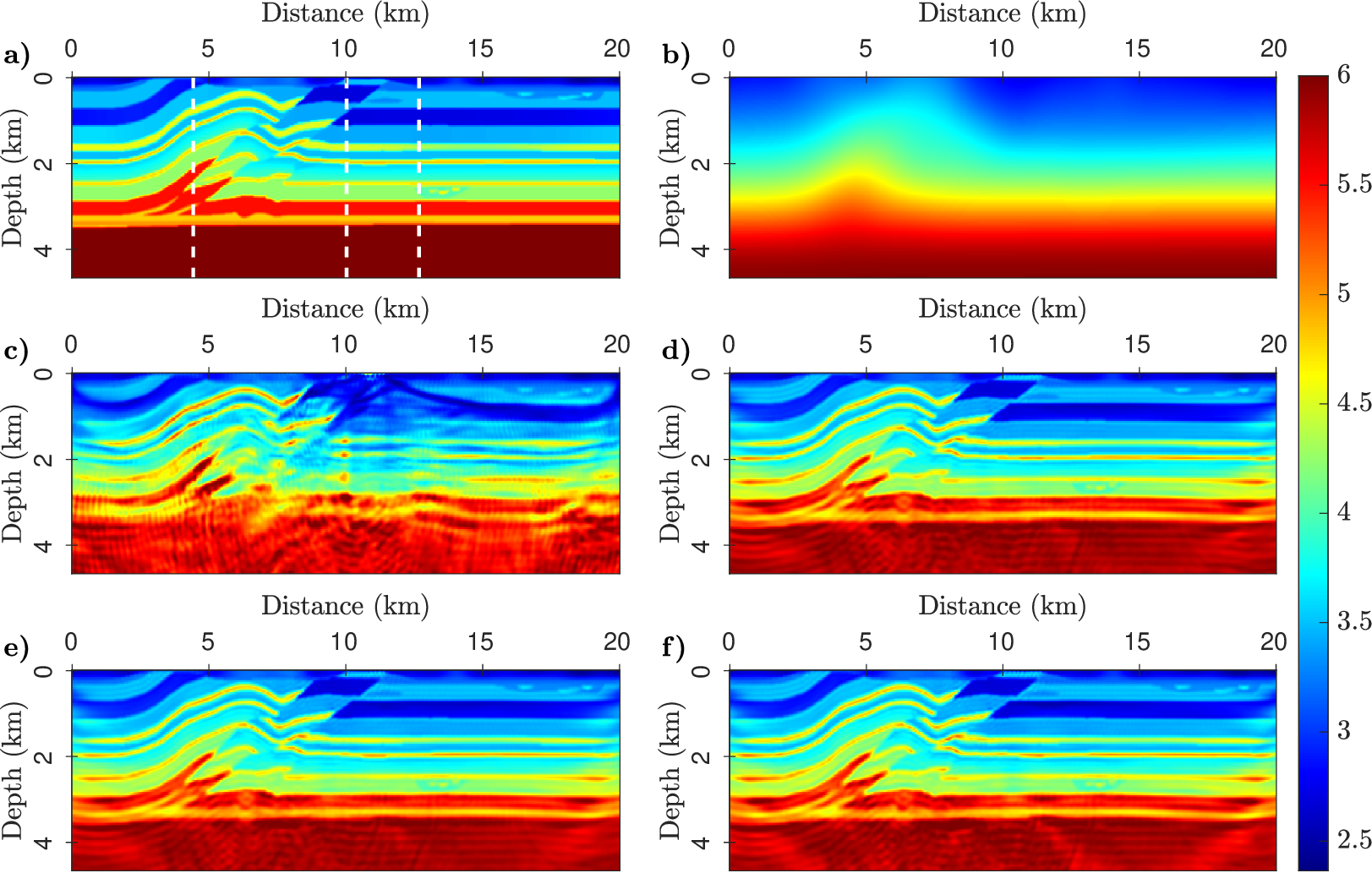}
\caption{The (a) Overthrust velocity model and (b) initial model. Inversion results obtained by solving the reduced FWI using (c) PSD, (d) EGN, (e) EGN only using $\bold{H}_r$, and (f) EGN only using $\bold{H}_s$.}
\label{Overthrust_estimates}
\end{figure}

\begin{figure}[!ht]
\center
\includegraphics[width=1\columnwidth]{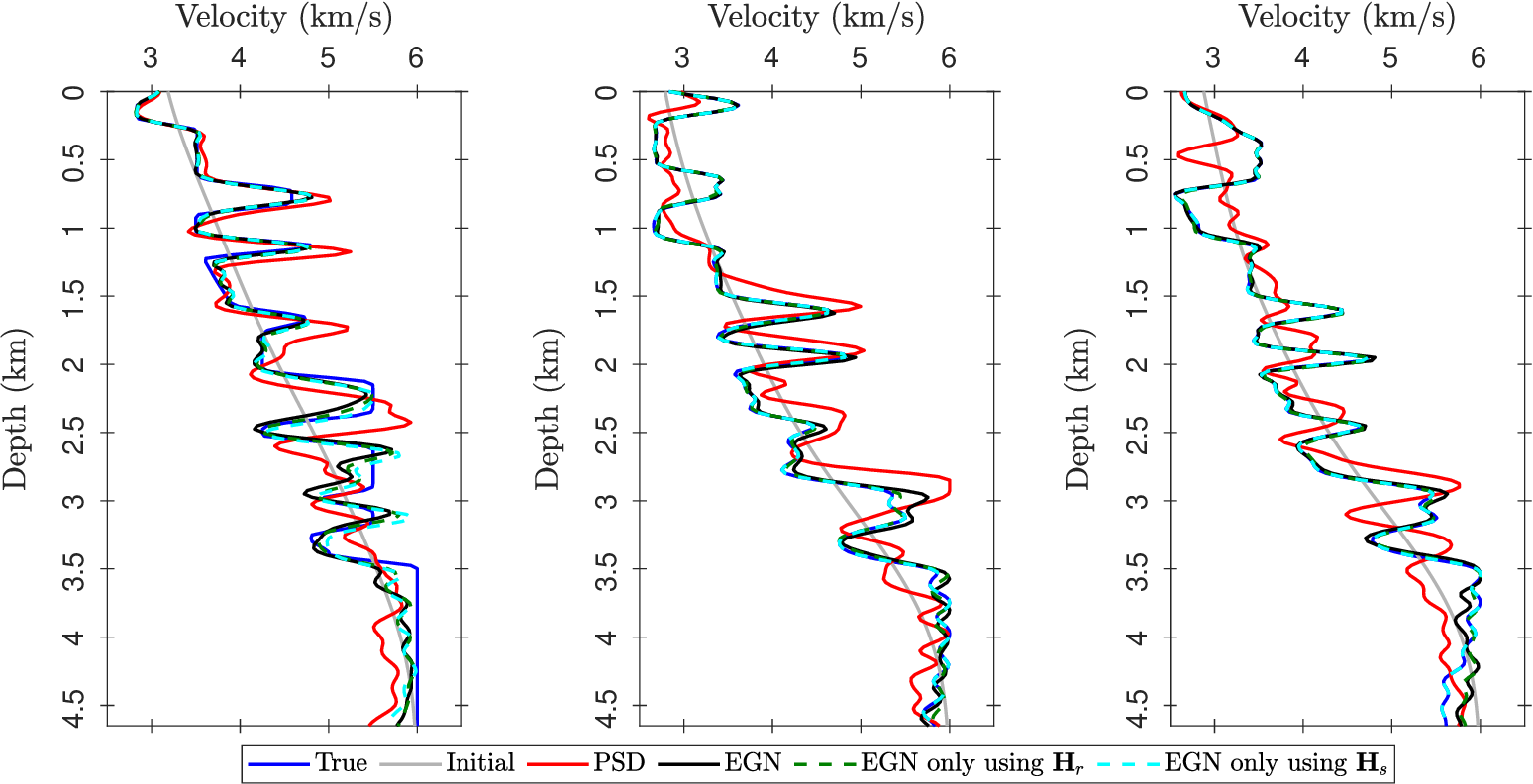}
\caption{Velocity logs extracted at locations indicated by white dashed lines on Figure \ref{Overthrust_estimates}a. The plot provides a direct comparison between the true model (blue line), the initial model (gray line), and the reconstructed models obtained solving the reduced FWI using the PSD (red line), EGN (black line), EGN only using $\bold{H}_r$ (green line), and EGN only using $\bold{H}_s$ (cyan line).}
\label{Overthrust_logs}
\end{figure}

\begin{figure}[!ht]
\center
\includegraphics[width=1\columnwidth]{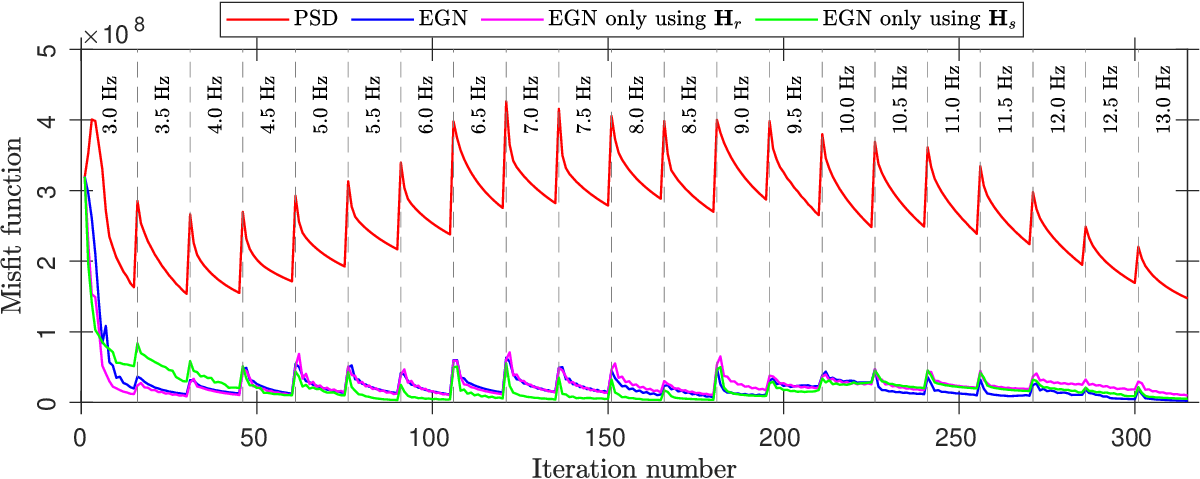}
\caption{Trajectory of the misfit function versus iteration for the Overthrust model obtained by solving the reduced FWI using the PSD (red line), EGN (blue line), EGN only using $\bold{H}_r$ (magenta line), and EGN only using $\bold{H}_s$ (green line).}
\label{Overthrust_conv}
\end{figure}

\noindent We assess the EGN method against the Overthrust model and follow a multiscale inversion approach. The subsurface model has dimensions of 20 km width and 4.6 km depth, discretized with a 25 m grid interval (Figure \ref{Overthrust_estimates}a). The fixed spread acquisition consists of 134 equally-spaced sources and 134 equally-spaced receivers on the top side of the model. We implement a PML absorbing boundary condition around the model and use a Ricker wavelet with a dominant frequency of 10 Hz as the source signature. We start the inversion from a strongly smoothed initial model (Figure \ref{Overthrust_estimates}b) and perform one cycle of inversion using multiscale frequency continuation from 3 Hz to 13 Hz with a 0.5 Hz frequency interval. Each frequency is inverted individually with 15 iterations per frequency, and the estimated model from each frequency is used as the starting model for the next frequency. The final results obtained by the PSD and EGN methods, solving the reduced FWI, are shown in Figures \ref{Overthrust_estimates}c and d, respectively. The results obtained using EGN for penalty FWI were similar to those obtained for reduced FWI with $\bold{h}=\bold{0}$ (Figure \ref{Overthrust_estimates}d).
Figure \ref{Overthrust_estimates}c demonstrates that the result obtained by the PSD method is inaccurate and includes artifacts, while the EGN result is accurate without artifacts.
A direct comparison among the true model, the initial model, and the estimated models along three vertical logs indicated by white dashed lines on Figure \ref{Overthrust_estimates}a is shown in Figure \ref{Overthrust_logs}.
Additionally, the associated convergence curves in Figure \ref{Overthrust_conv} show that EGN exhibits much faster convergence compared to PSD.

Furthermore, we examined the effect of each Hessian matrix $\bold{H}_s$ and $\bold{H}_r$ separately. Figure \ref{Overthrust_estimates}e shows the result obtained by EGN using only $\bold{H}_r$, while Figure \ref{Overthrust_estimates}f is obtained using only $\bold{H}_s$. In this example, since the same number of sources and receivers were used at the surface, both $\bold{H}_s$ and $\bold{H}_r$ have a similar effect and generate high-resolution models. However, from the associated convergence curves (Figure \ref{Overthrust_conv}), we observe that $\bold{H}_r$ better fits low frequencies, while $\bold{H}_s$ achieves a lower misfit at high frequencies. When both matrices are used by EGN, an improved misfit is obtained at both low and high frequencies.
Moreover, we find that $\bold{H}_s$ may be a better alternative to the traditional pseudo-Hessian matrix \citep{Shin_2001_IAP}. $\bold{H}_s$ is formed by $\S\S^T$ and properly removes the effect of the Hessian from the source dimension of the data residuals, while the pseudo-Hessian matrix uses the diagonal of this matrix and simply scales the gradient. By comparing Figures \ref{Overthrust_estimates}c and f, we can see that $\bold{H}_s$ is more effective.


\subsection{Examples with randomized EGN}
\begin{figure}
\center
\includegraphics[width=1\columnwidth]{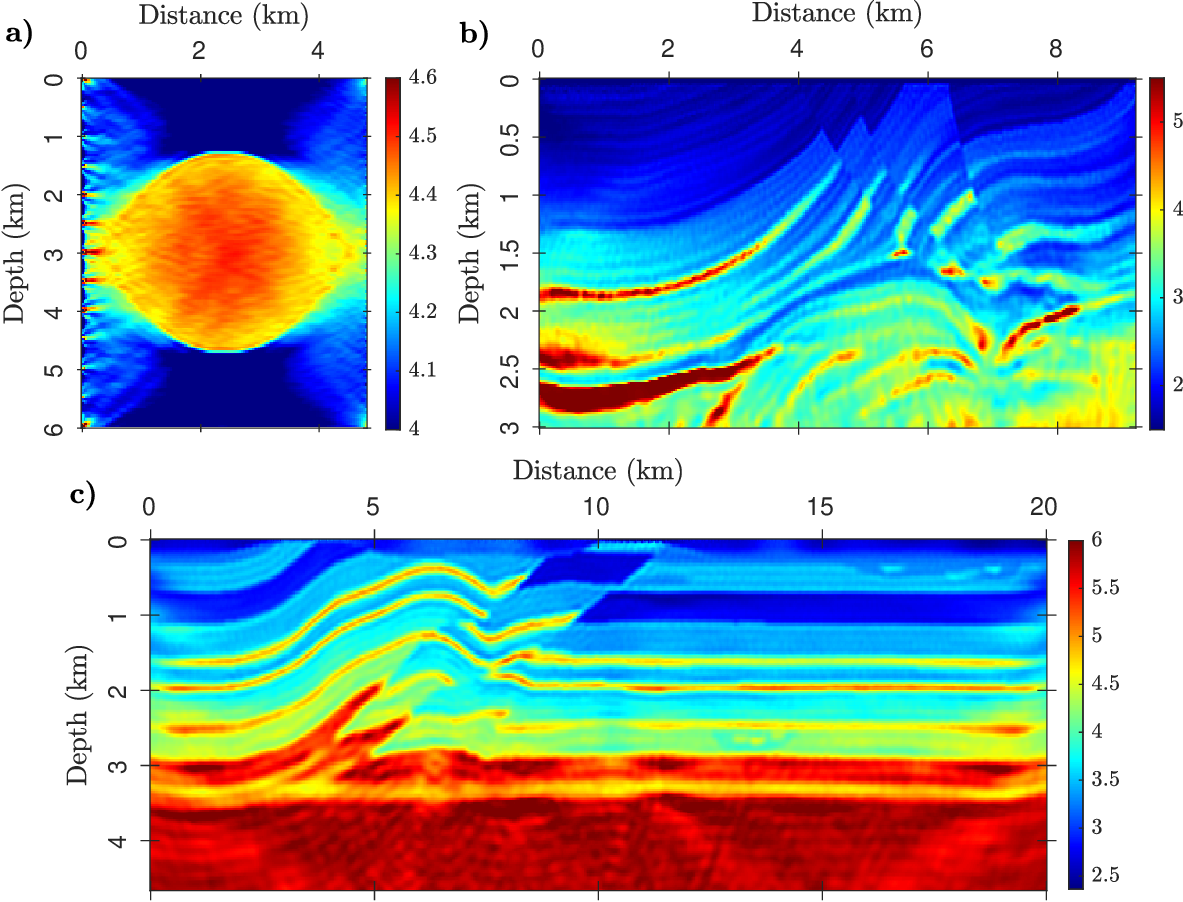}
\caption{Inversion results obtained by the randomized EGN method solving the reduced FWI. The (a) Camembert model, (b) Marmousi model, and (c) Overthrust model.}
\label{Randomized}
\end{figure}

\noindent Finally, using the same parameters as the previous examples, we assess the performance of the EGN method against source-receiver sketching. We employ the simplest form of EGN that solves the reduced FWI with $\bold{h=0}$. The randomized EGN reduces to the deterministic algorithm when the sketching matrices are identity matrices. We employ sketching matrices with random Gaussian entries. For the Camembert and Marmousi examples, where multiple frequencies are inverted at each iteration, we set $N_p=N_q=10$. For the Overthrust example, where a single frequency is inverted at each iteration, we set $N_p=N_q=30$. The final results obtained by the randomized EGN method are shown in Figure \ref{Randomized}. By comparing Figures \ref{Randomized}(a-c) with the associated deterministic results in Figures \ref{Camembert}(e), \ref{Marmousi_I}(c), and \ref{Overthrust_estimates}(d), we observe that the randomized EGN method still produces high-resolution velocity models while reducing the overall number of required wave-equation solves by a factor of 9, 10, and 4, respectively, compared to their deterministic counterparts.


\section{Discussions}
\noindent The proposed EGN method offers a novel approach to FWI, expanding upon the traditional GN method by extending the search direction along the subsurface offset. This extension allows for a more comprehensive exploration of the model space, resulting in improved inversion results.
By incorporating all possible subsurface offsets in the range and direction \citep{Claerbout_1985_IEI,Symes_2008_MVA}, our method decomposes the GN Hessian into separate Hessian matrices, which can be efficiently applied along the source and receiver dimensions of the data residual matrix. This decomposition enhances the computational efficiency of the algorithm.
For the case of simultaneous frequency inversion, where multiple frequencies are inverted at each iteration, we tackle the challenge posed by the generalized Sylvester equation \ref{GSE} by inverting each frequency separately and subsequently stacking the results over frequency. This strategy effectively extends the search direction along the frequency axis, providing an additional level of exploration in the model space.

The extension of the EGN method to the time-domain formulation of FWI involves extending the search direction in both the subsurface offset and propagation time axes, resulting in equations similar to those obtained in the frequency-domain formulation. In the time-domain EGN formulation, the data residual, denoted as $\Delta\bold{d}$, remains a matrix, where each column represents a shot gather reshaped as a vector. Consequently, the Hessian matrix $\G\G^T$ takes the form of a block matrix with dimensions $N_rN_t\times N_rN_t$, consisting of $N_r\times N_r$ blocks, each of size $N_t\times N_t$, as illustrated in Figure A-1 of \citep{Gholami_2022_EFW}. The source-side Hessian matrix, $\S\S^T$, retains its size of $N_s\times N_s$.

The primary challenge encountered in the time-domain implementation of the EGN method is related to the construction and inversion of the  $\G\G^T$ matrix. To address this, a truncated EGN approach can be employed to estimate the extended data residuals by solving the corresponding linear system using the CG iteration. This truncated EGN method requires two additional wave-equation solves per CG iteration. However, it is worth noting that this extra computational cost is also present when applying the truncated GN method with the second-order adjoint state algorithm for FWI \citep{Metivier_2017_TRU}.
Consequently, the computational cost of the EGN method using the matrix-free formalism is comparable to that of the standard GN method. Once the extended data residual matrix is obtained, the subsequent steps involved in the EGN algorithm remain the same as in the standard FWI algorithm. These steps include the multiplication of the forward and backward wavefields and the summation over time and source.  Summation over the subsurface offset can also be done on the fly by shifting the propagating wavefields in space.
By utilizing the truncated EGN approach and the matrix-free formalism, we can effectively address the challenge associated with constructing and inverting the $\G\G^T$ matrix in the time-domain implementation of the EGN method. This enables us to achieve computational efficiency comparable to the standard GN method while benefiting from the enhanced capabilities of the EGN method.

Finally, our investigations on various numerical examples have revealed that the impact of model extension on improving inversion results is significantly more pronounced compared to the effect of source extension. Although the source extension plays a preconditioning role in the EGN framework (compare, e.g., Figure \ref{Camembert}f with the reference \ref{Camembert}e and Figure \ref{Camembert_dm}d with the reference \ref{Camembert_dm}c), further investigation is needed to determine the specific conditions under which the source extension becomes more significant and influential. Understanding these conditions will provide valuable insights into optimizing the usage of source extension within the EGN method.

\section{Conclusions}
\noindent We have developed the Extended Gauss-Newton (EGN) method for multisource multireceiver waveform inversion, which extends the search direction of the standard Gauss-Newton (GN) method. After rewriting the GN system as an equivalent matrix equation with a diagonal matrix as the solution instead of a vector, we demonstrated how relaxing the diagonal constraint added a degree of freedom for search direction along the subsurface offset axis. A separable Hessian is produced as a result of this relaxation of the diagonal constraint, making efficient inversion possible. The EGN method offers an alternative approach to extended FWI that allows us to use the advantages of both model extension and source extension methods, while maintaining computational efficiency of the reduced FWI. The efficiency and stability of EGN in estimating high-quality velocity models have been illustrated by several numerical examples. It displayed strong performance, consistently delivering accurate inversion results. The method is a useful tool for tackling challenging FWI problems because of its efficiency and robustness.

\section{ACKNOWLEDGMENTS}  
This research was financially support by the SONATA BIS grant (No. 2022/46/E/ST10/00266) of the National Science Center in Poland. 

\newcommand{\SortNoop}[1]{}


\begin{thebibliography}{}
\itemsep0pt

\bibitem[Abubakar et~al., 2009]{Abubakar_2009_FDC}
Abubakar, A., W. Hu, T.~M. Habashy, and P.~M. {van den Berg},  2009,
  Application of the finite-difference contrast-source inversion algorithm to
  seismic full-waveform data: Geophysics, {\bf 74}, WCC47--WCC58.

\bibitem[Aghazade et~al., 2022]{Aghazade_2021_RSS}
Aghazade, K., A. Gholami, H. Aghamiry, and S. Operto,  2022, Randomized source
  sketching for full waveform inversion: {IEEE} {T}ransactions on {G}eoscience
  and {R}emote {S}ensing, {\bf 60}, 1--12.

\bibitem[Biondi and Almomin, 2014]{Biondi_2014_SIF}
Biondi, B. and A. Almomin,  2014, Simultaneous inversion of full data bandwidth
  by tomographic full-waveform inversion: Geophysics, {\bf 79(3)},
  WA129--WA140.

\bibitem[Biondi and Shan, 2002]{Biondi_2002_PIO}
Biondi, B. and G. Shan,  2002, Prestack imaging of overturned reflections by
  reverse time migration, {\it in} SEG Technical Program Expanded Abstracts
  2002,  1284--1287. Society of Exploration Geophysicists.

\bibitem[Biondi and Symes, 2004]{Biondi_2004_ADC}
Biondi, B. and W. Symes,  2004, Angle-domain common-image gathers for migration
  velocity analysis by wavefield-continuation imaging: Geophysics, {\bf 69(5)},
  1283--1298.

\bibitem[Claerbout, 1985]{Claerbout_1985_IEI}
Claerbout, J.,  1985, Imaging the {E}arth's interior: {Blackwell} Scientific
  Publication.

\bibitem[Fu and Symes, 2017]{Fu_2017_AMA}
Fu, L. and W.~W. Symes,  2017, An adaptive multiscale algorithm for efficient
  extended waveform inversion: Geophysics, {\bf 82}, R183--R197.

\bibitem[Gauthier et~al., 1986]{Gauthier_1986_TDN}
Gauthier, O., J. Virieux, and A. Tarantola,  1986, Two-dimensional nonlinear
  inversion of seismic waveforms: numerical results: Geophysics, {\bf 51},
  1387--1403.

\bibitem[Gholami et~al., 2022a]{Gholami_2022_EFW}
Gholami, A., H.~S. Aghamiry, and S. Operto,  2022a, Extended full waveform
  inversion in the time domain by the augmented {L}agrangian method:
  Geophysics, {\bf 87}, R63--R77.

\bibitem[Gholami et~al., 2022b]{Gholami_2022_OCN}
--------, 2022b, On the connection between {WRI} and {FWI}: Analysis of the
  nonlinear term in the hessian matrix, {\it in} SEG Technical Program Expanded
  Abstracts 2022,  1--5. Society of Exploration Geophysicists.

\bibitem[Guttman, 1946]{Guttman_1946_EMC}
Guttman, L.,  1946, Enlargement methods for computing the inverse matrix: The
  annals of mathematical statistics,  336--343.

\bibitem[Hautus, 1994]{Hautus_1994_OS}
Hautus, M. L.~J.,  1994, Operator substitution: Linear algebra and its
  applications, {\bf 205}, 713--739.

\bibitem[Horn and Johnson, 1994]{Horn_1994_TMA}
Horn, R.~A. and C.~R. Johnson,  1994, Topics in matrix analysis: Cambridge
  University Press.

\bibitem[Hou and Symes, 2015]{Hou_2015_AIE}
Hou, J. and W.~W. Symes,  2015, An approximate inverse to the extended {B}orn
  modeling operator: Geophysics, {\bf 80}, R331--R349.

\bibitem[Huang et~al., 2018]{Huang_2018_VSE}
Huang, G., R. Nammour, and W.~W. Symes,  2018, Volume source-based extended
  waveform inversion: Geophysics, {\bf 83}, R369--387.

\bibitem[Lindsey, 1989]{Lindsey_1989_FZI}
Lindsey, J.~P.,  1989, The {Fresnel} zone and its interpretative significance:
  The Leading Edge, {\bf 8}, 33--39.

\bibitem[M{\'e}tivier et~al., 2017]{Metivier_2017_TRU}
M{\'e}tivier, L., R. Brossier, S. Operto, and V. J.,  2017, Full waveform
  inversion and the truncated {N}ewton method: SIAM Review, {\bf 59}, 153--195.

\bibitem[Plessix, 2006]{Plessix_2006_RAS}
Plessix, R.~E.,  2006, A review of the adjoint-state method for computing the
  gradient of a functional with geophysical applications: Geophysical Journal
  International, {\bf 167}, 495--503.

\bibitem[Pratt et~al., 1998]{Pratt_1998_GNF}
Pratt, R.~G., C. Shin, and G.~J. Hicks,  1998, {G}auss-{N}ewton and full
  {N}ewton methods in frequency-space seismic waveform inversion: Geophysical
  Journal International, {\bf 133}, 341--362.

\bibitem[Rickett and Sava, 2002]{Rickett_2002_OAC}
Rickett, J. and P. Sava,  2002, Offset and angle-domain common image-point
  gathers for shot-profile migration: Geophysics, {\bf 67}, 883--889.

\bibitem[Sava and Fomel, 2003]{Sava_2003_ADC}
Sava, P.~C. and S. Fomel,  2003, Angle-domain common-image gathers by wavefield
  continuation methods: Geophysics, {\bf 68}, 1065--1074.

\bibitem[Sheriff, 1980]{Sheriff_1980_NFC}
Sheriff, R.~E.,  1980, Nomogram for {Fresnel}-zone calculation: Geophysics,
  {\bf 45}, 968--972.

\bibitem[Shin et~al., 2001]{Shin_2001_IAP}
Shin, C., S. Jang, and D.~J. Min,  2001, Improved amplitude preservation for
  prestack depth migration by inverse scattering theory: Geophysical
  Prospecting, {\bf 49}, 592--606.

\bibitem[Symes, 2008]{Symes_2008_MVA}
Symes, W.~W.,  2008, Migration velocity analysis and waveform inversion:
  Geophysical Prospecting, {\bf 56}, 765--790.

\bibitem[Symes, 2020]{Symes_2020_WRI}
--------, 2020, Wavefield reconstruction inversion: an example: Inverse
  Problems, {\bf 36}, 105010.

\bibitem[Symes and Kern, 1994]{Symes_1994_IRSb}
Symes, W.~W. and M. Kern,  1994, Inversion of reflection seismograms by
  differential semblance analysis: algorithm structure and synthetic examples:
  Geophysical Prospecting, {\bf 42}, 565--614.

\bibitem[Tarantola, 1984]{Tarantola_1984_ISR}
Tarantola, A.,  1984, Inversion of seismic reflection data in the acoustic
  approximation: Geophysics, {\bf 49}, 1259--1266.

\bibitem[van Leeuwen, 2019]{vanLeeuwen_2019_ANO}
van Leeuwen, T.,  2019, A note on extended full waveform inversion: arXiv
  preprint arXiv:1904.00363.

\bibitem[{van Leeuwen} and Herrmann, 2016]{vanLeeuwen_2016_PMP}
{van Leeuwen}, T. and F. Herrmann,  2016, A penalty method for
  {PDE}-constrained optimization in inverse problems: Inverse Problems, {\bf
  32(1)}, 1--26.

\bibitem[{van Leeuwen} and Herrmann, 2013]{VanLeeuwen_2013_MLM}
{van Leeuwen}, T. and F.~J. Herrmann,  2013, Mitigating local minima in
  full-waveform inversion by expanding the search space: Geophysical Journal
  International, {\bf 195(1)}, 661--667.

\bibitem[Virieux and Operto, 2009]{Virieux_2009_OFW}
Virieux, J. and S. Operto,  2009, An overview of full waveform inversion in
  exploration geophysics: Geophysics, {\bf 74}, WCC1--WCC26.

\bibitem[Wang et~al., 2017]{Wang_2017_SMR}
Wang, J., J. Lee, M. Mahdavi, M. Kolar, and N. Srebro,  2017, Sketching meets
  random projection in the dual: A provable recovery algorithm for big and
  high-dimensional data: Electronic Journal of Statistics, {\bf 11},
  4896--4944.

\bibitem[Yang et~al., 2019]{Yang_2019_LSR}
Yang, J., Y. Elita~Li, A. Cheng, Y. Liu, and L. Dong,  2019, Least-squares
  reverse time migration in the presence of velocity errors: Geophysics, {\bf
  84}, S567--S580.

\end{thebibliography}
\end{document}